\documentclass[11pt]{article}
\usepackage{amsfonts,amssymb}
\usepackage{color}
\usepackage{pstricks,pst-node}
\usepackage{pst-plot}
\usepackage{mathtools}
\catcode`\@=11
\def\marginnote#1{}

\newcount\hour
\newcount\minute
\newtoks\amorpm
\hour=\time\divide\hour by60 \minute=\time{\multiply\hour by60
\global\advance\minute by-\hour}
\edef\standardtime{{\ifnum\hour<12 \global\amorpm={am}%
        \else\global\amorpm={pm}\advance\hour by-12 \fi
        \ifnum\hour=0 \hour=12 \fi
        \number\hour:\ifnum\minute<10 0\fi\number\minute\the\amorpm}}
\edef\militarytime{\number\hour:\ifnum\minute<10 0\fi\number\minute}

\def\appendix#1{
\addtocounter{section}{1} \setcounter{equation}{0}
\renewcommand{\thesection}{\Alph{section}}
\section*{Appendix \thesection\protect\indent\quad
#1}
}
\renewcommand{\theequation}{\thesection.\arabic{equation}}

\def\draftlabel#1{{\@bsphack\if@filesw {\let\thepage\relax
      \xdef\@gtempa{\write\@auxout{\string
          \newlabel{#1}{{\@currentlabel}{\thepage}}}}}\@gtempa \if@nobreak
    \ifvmode\nobreak\fi\fi\fi\@esphack} \gdef\@eqnlabel{#1}}
    \def\@eqnlabel{}
\def\@vacuum{}
\def\draftmarginnote#1{\marginpar{\raggedright\scriptsize\tt#1}}

\def\draft{
  \oddsidemargin -.5truein
  \def\@oddfoot{\footnotesize \sl preliminary draft \hfil
    \rm\thepage\hfil\sl\today\quad\militarytime}
  \let\@evenfoot\@oddfoot \overfullrule 3pt
    \let\label=\draftlabel
    \let\marginnote=\draftmarginnote
  \def\@eqnnum{(\theequation)\rlap{\kern\marginparsep\tt\@eqnlabel}%
    \global\let\@eqnlabel\@vacuum}

  }

\voffset= - 0.5in \hoffset= - 1.0in

\def\be{\begin{equation}}
\def\ee{\end{equation}}
\def\bea{\begin{eqnarray}}
\def\eea{\end{eqnarray}}
\def\<{\langle}
\def\>{\rangle}

\def\res{{{\rm res}}}
\def\Im{{\rm Im}}

\def\tr{{\mathrm{tr\,}}}

\def\1N{${\cal N}=1$}
\def\4N{${\cal N}=4$}

\newcommand{\bp}{\mathbb{P}}
\newcommand{\bz}{\mathbb{Z}}
\newcommand{\modm}{\mathcal{M}}

\def\e{{\,\rm e}\,}

\def\bea{\begin{eqnarray}}
\def\eea{\end{eqnarray}}

\def\pa{\partial}

\def\beq{\begin{equation}}
\def\eeq{\end{equation}}
\def\ba{\beq\begin{array}{c}}
\def\ea{\end{array}\eeq}

\@ifundefined{theorem@style}{\RequirePackage{theorem}}{}

\gdef\th@plain{\normalfont\slshape
  \def\@begintheorem##1##2{%
\item[\hskip\parindent\hskip\labelsep\theorem@headerfont ##1\ ##2\unskip.]}%
\def\@opargbegintheorem##1##2##3{%
\item[\hskip\parindent
\ifx\empty##1\else\hskip\labelsep\fi\theorem@headerfont ##1\ ##2\unskip]{\theorem@headerfont{\rm ##3}.} }}

\gdef\th@definition{\normalfont
  \def\@begintheorem##1##2{%
\item[\hskip\parindent\hskip\labelsep\theorem@headerfont ##1\ ##2\unskip.]}%
\def\@opargbegintheorem##1##2##3{%
\item[\hskip\parindent
\ifx\empty##1\else\hskip\labelsep\fi\theorem@headerfont ##1\ ##2\unskip]{\theorem@headerfont{\rm ##3}.} }}

\theoremstyle{plain}
\newtheorem{theorem}{Theorem}
\newtheorem{lemma}{Lemma}
\newtheorem{corollary}{Corollary}
\newtheorem{proposition}{Proposition}

\theoremstyle{definition}

\newtheorem{definition}{Definition}
\newtheorem{remark}{Remark}
\newtheorem{example}{Example}

\let\operatorname=\mathrm
\let\text=\mathrm

\def\ln{\operatorname{log}}

\def\Aut{\operatorname{Aut}}

\def\disc{\text{disc}}

\def\beq{\begin{equation}}
\def\eeq{\end{equation}}
\def\bea{\begin{eqnarray}}
\def\eea{\end{eqnarray}}

\renewcommand{\binom}[2]{\left({#1\atop #2}\right)}

\newcommand{\cpict}[3]{
\dimen1=#1\advance\dimen1 by-\hsize\divide\dimen1 by-2 \vtop to #2{
\noindent\hskip\dimen1{\special{em:graph #3.bmp}} \vfil}\hskip-2cm }

\let\@@savethanks\thanks
\def\thanks#1{\gdef\thefootnote{\alph{footnote}}\@@savethanks{#1}}

\newcommand{\cl}{\mathcal{L}}
\newcommand{\bc}{\mathbb{C}}

\newcommand {\dvol}{discrete volumes}

\makeatletter
\def\blfootnote{\xdef\@thefnmark{}\@footnotetext}
\makeatother

\begin{document}

\title{Topological recursion for Gaussian means and cohomological field theories}

\author{J{\o}rgen Ellegaard Andersen\thanks{\noindent QGM, {\AA}rhus University, Denmark and Caltech, Pasadena, USA},
Leonid O. Chekhov\thanks{Steklov Mathematical Institute of Russian Academy of Sciences, Moscow, Russia},
Paul Norbury\thanks{University of Melbourne, Australia}, and
Robert C. Penner\thanks{IHES, Bures-sur-Yvette, France, and Caltech, Pasadena, USA}}

\date{}


\maketitle

\begin{abstract}
We use the explicit relation between genus filtrated $s$-loop means 
of the Gaussian matrix model and terms of the genus expansion of the 
Kontsevich--Penner matrix model (KPMM), which is the
generating function for volumes of discretized (open) moduli spaces
$M_{g,s}^{\disc}$ (discrete volumes), to express Gaussian means in all genera as polynomials
in special times weighted by ancestor invariants of an underlying cohomological field theory.
We translate topological recursion of the Gaussian model into recurrent relations for coefficients of this expansion
proving their integrality and positivity. 
As an application, we find the coefficients in the first subleading order for ${\mathcal M}_{g,1}$ for 
all $g$ in three ways:
by using the refined Harer--Zagier recursion, by exploiting the Givental-type decomposition of KPMM, and
by an explicit diagram counting.
\end{abstract}

\section{Introduction}
\setcounter{equation}{0}

Multi-trace means $\<\prod_{i=1}^s \tr H^{k_i}\>^{{\mathrm{conn}}}$ of the Gaussian Unitary Ensemble (GUE) were under investigation for many years. First, Harer and Zagier obtained~\cite{HZ} the linear recursion formula on genus filtrated one-trace means, which allows 
obtaining answers for very high genera (unattainable by other tools). Although exact  
$s$-fold integral representation for $s$-trace means valid for all $N$ were obtained by Brez\'in and Hikami~\cite{BreHik} using the replica
method ameliorated in~\cite{MS10}, producing an effective genus expansion on the base of these formulas still remains an open problem.
The interest to multi-trace means was revived after the appearance of {\em topological recursion} \cite{ChEy,EOrInv} and quantum curves \cite{GSuApo,MSuSpe,DM14}.  It was shown in our first paper \cite{ACNP} that Gaussian means are related via the so-called Kontsevich--Penner
matrix model (KPMM)~\cite{ChMak1,ACKM}
to discrete volumes of open moduli spaces and, simultaneously, to generating functions of ancestor invariants of a
{\em cohomological field theory} \cite{ManFro}.

We come to the KPMM using explicit combinatorial formulas.  
It is known since~\cite{ChMak2} and~\cite{MMM} that the KPMM is
equivalent to the Hermitian matrix model with the potential whose times (coupling constants) are related to
the external-matrix eigenvalues via the Miwa-type transformation and whose matrix size is the coefficient of
the logarithmic term.  The first result of \cite{ACNP} is that the KPMM is a primitive (antiderivative) for the resolvents of the Gaussian matrix model.  The resolvents storing the multi-trace Gaussian means are naturally described as meromorphic (multi)differentials with zero residues over a rational Riemann surface, known as the {\em spectral curve}, hence their primitives are meromorphic functions on the spectral curve.  These primitives are conjecturally related (this was proven in the Gaussian case \cite{MSuSpe}, see also \cite{Norbury15}) to the so called {\em quantum curve} which is a linear differential equation that is a non-commutative quantisation of the spectral curve.  
The spectral and quantum curves are related: the wave function 
emerging out of the spectral curve is a specialization of the free energy for the KPMM which satisfies the second order differential equation that is the quantum curve.

The geometric content of the KPMM is also rich: its free energy was related to structures of {\em discretized moduli spaces} in~\cite{Ch93} and it was identified recently (see~\cite{NorStr} and \cite{Mulase-Penkawa}) with the generating function for \dvol\ 
$N_{g,s}(P_1,\dots,P_s)$---quasi-polynomials introduced in~\cite{Norbury} that
count integer points in the interiors ${\mathcal M}_{g,s}$ of moduli spaces of Riemann surfaces of genus $g$ with $s>0$ holes with the
fixed perimeters $P_j\in{\mathbb Z}_+$, $j=1,\dots,s$ of holes in the Strebel uniformization.
Moreover, it was shown in~\cite{Ch95} that in the special times $T^{\pm}_{2n}$ that are discrete Laplace transforms of monomials
$P_I^{2k}$, this model admits a decomposition into two Kontsevich models related by a Bogolyubov canonical
transformation, which was the first example of the Givental-type
decomposition formulas~\cite{Giv}. We use the approach of~\cite{Ch95}  for presenting the
free-energy expansion terms ${\mathcal F}_{g,s}$ of the KPMM as finite sums over graphs whose nodes are
terms of the expansion of the Kontsevich
matrix model free energy, internal edges correspond to quadratic terms in the canonical transformation operator, external
half edges (dilaton leaves) correspond to the constant shifts of the higher times, and external legs
(ordinary leaves) carry the times $T^{\pm}_{2n}$.
This graph representation provides another proof of quasi-polynomiality of $N_{g,s}(P_1,\dots,P_s)$.

From \cite{Ey11} and \cite{DOSSIde} we know that the terms of topological recursion \cite{Ey},\cite{ChEy},\cite{CEO},\cite{AlMM} 
based on a certain
spectral curve satisfying a compatibility condition (relating the $w_{0,1}$ and $w_{0,2}$ invariants) describe ancestor invariants of a cohomological field theory (CohFT), or equivalently a Frobenius manifold.

A fundamental family of Frobenius manifolds described by Dubrovin are Hurwitz spaces.  For $\mu=(\mu_1,\dots,\mu_n)$, 
the Hurwitz space $H_{g,\mu}$  consists of homotopy classes of
genus $g$ branched covers of the sphere with $n$ labeled points over $\infty$ of ramification profile $(\mu_1,\dots,\mu_n)$ and simple ramification over $\bp^1-\infty$.  It has dimension $|\mu|+n+2g-2$ where $|\mu|=\mu_1+\cdots+\mu_n$.

The 2-dimensional Hurwitz--Frobenius manifold $H_{0,(1,1)}$ consists of double branched covers of the sphere with two branch points and no ramification at infinity.  Its free energy is
\beq
F_0(t_{0,1},t_{0,0})=\frac{1}{2}t_{0,0}^2t_{0,1}+\frac{1}{2}t_{0,1}^2\log{t_{0,1}}-\frac34 t^2_{0,1}
\label{Frob}
\eeq
with the Euler vector field $E=t_{0,0}\frac{\partial}{\partial t_{0,0}}+2t_{0,1}\frac{\partial}{\partial t_{0,1}}.$
Note that expression (\ref{Frob}) appears as a standard term (the perturbative part) in the
expansion of any matrix model upon identification of $t_{0,1}$ with the normalized number of eigenvalues and $t_{0,0}$
with the first time; we have that
$$
\log\int \prod_{i=1}^{t_{0,1}N}dx_i\,\prod_{i<j}(x_i-x_j)^2 \,e^{-N\sum_{i=1}^{t_{0,1}N} \bigl(\frac12 x_i^2-t_{0,0}x_i\bigr)}
=N^2 F_0(t_{0,1},t_{0,0})+\sum_{g=1}^\infty N^{2-2g} F_g(t_{0,1}),
$$
where the leading term of the $1/N$-expansion of the free energy of the above Gaussian matrix model is exactly (\ref{Frob}).

In \cite{ACNP}, we related the \dvol\  to the Gaussian means $W^{(g)}_{s}(x_1,\dots,x_s)$
and used the CohFT description further relating
the \dvol\  to ancestor invariants of a CohFT. These ancestor invariants are evaluated already
in terms of the {\em closed} moduli spaces ${\mathcal M}_{g,s}$ compactified by Deligne and Mumford.

The paper is organized as follows. In Sec.~\ref{s:matrix-model} we establish the equivalence between the Gaussian
means (the correlation functions) and the terms of expansion of the KPMM free energy. 

In Sec.~\ref{s:KPMM}, we describe the results of~\cite{Ch93}, \cite{Ch95}, and \cite{Norbury} for open discrete
moduli spaces, which we used in \cite{ACNP} to relate the above Gaussian means and the \dvol\  in a purely combinatorial way. 
The quantum curve can then be obtained as a specialization of the KPMM to the case of unit size matrices. We describe the Givental-type
decomposition formulas for the KPMM obtained in \cite{Ch95} representing them in terms of graph expansions for
the free energy terms. This graph representation also implies the quasi-polynomiality of
the \dvol\  and provides a link to a CohFT.  

In Sec.~\ref{s:CohFT}, we identify the Gaussian means expansion terms with the ancestor invariants of
a cohomological field theory using the results of \cite{DNOPSSup} and \cite{DOSSIde}. The decomposition thus obtained has a canonical Givental form. The coefficients of 
this decomposition, or Laplace transforms of the quasi-polynomials $N_{g,s}(P_1,\dots,P_s)$,
are the special coefficients $\widehat b_{\vec k,\vec \beta}^{(g)}$, 
which in a sense represent in the ``most economic'' way the
genus filtered $s$-loop means $W_s^{(g)}(x_1,\dots, x_s)$ and 
are linear combinations of the CohFT ancestor invariants of neighbouring levels. 

In Sec.~\ref{s:TR}, we develop the topological recursion for Gaussian means, present the general recursion relations for
$\widehat b_{\vec k,\vec \beta}^{(g)}$, and prove that in the range of admissibility all these coefficients are positive integers. 

In Sec.~\ref{s:HZ}, we concentrate on the case of
a one-loop mean. We find the first subleading coefficient $b^{(g)}_{g-2}$ in three
ways: using the modified Harer--Zagier (HZ) recurrence relation, by the graph description of Givental-type decomposition
in Sec.\ref{s:KPMM}, and
by an explicit diagram counting.

\section{The effective matrix model for the multi-loop Gaussian means}
\label{s:matrix-model}
\setcounter{equation}{0}

We consider a sum of connected \emph{chord diagrams} based on $s$ backbones, or loop insertions, carrying the
variables $u_i$, $i=1,\dots,s$. We first provide an effective matrix model description for all
genus-$g$ contributions in terms of {\em shapes} ---the connected fatgraphs of genus $g$ with $s$ faces and with vertices
of arbitrary order greater or equal three; from the Euler characteristic formula, for a fixed $g$ and $s$, 
only a finite number of such fat graphs exist, and we let $\Gamma_{g,s}$ denote this finite set.  This set $\Gamma_{g,s}$
enumerates cells in the canonical Strebel--Penner ideal cell decomposition of moduli space ${\mathcal M}_{g,s}$.
In accordance with \cite{APRW}, 
$\Gamma_{g,s}$ is in bijection with circular chord diagrams which are also "shapes" in the terminology of \cite{APRW}, that is chord diagrams which are seeds and which has no one-chords.

The correlation functions, or means, are given by the integrals
\be
\left\langle\prod_{i=1}^s(\tr H^{k_i})\right\rangle = \int_{H\in \mathcal{H}_{N}} \left(\prod_{i=1}^s\tr H^{k_i}\right) e^{-\frac N2 \tr H^2} DH,
\label{corr}
\ee
where $\mathcal{H}_{N}$ is the set of Hermitian $N\times N$ matrices. 
By Wick's theorem, any correlation function (\ref{corr}) can be presented as the sum over all possible (complete) pairings between matrix entries $M_{ij}$, where the pairings are two-point correlation functions
$\langle H_{i,j}H_{k,l}\rangle =\frac 1N \delta_{il}\delta_{jk}$. These pairing are customarily represented
by \emph{edges}: double lines of indices. The corresponding index lines
run along faces of  \emph{fatgraphs} containing ordered set of $s$ vertices of valencies $k_i$, $i=1,\dots,s$, and $\sum_{i=1}^s k_i/2$ edges. 
For each vertex, we fix a cyclic order of edges incident to this vertex.
Furthermore for each vertex we also have a first incident edge given. We denote this set of fatgraphs $\widehat\Gamma(k_1,\dots,k_s)$. Then the sum in (\ref{corr}) becomes
$\sum_{\gamma\in \widehat\Gamma(k_1,\dots,k_s)}
N^{b(\gamma)-\sum_{i=1}^s k_i/2}$, where $b(\gamma)$ is the number of boundary components of $\gamma$.

Let $\widehat\Gamma(k_1,\dots,k_s)^c$ be the subset of $\widehat\Gamma(k_1,\dots,k_s)$ which consist of connected fatgraphs and let
$\Bigl\langle\prod_{i=1}^s(\tr H^{k_i})\Bigr\rangle^{\mathrm{conn}}$ be the part of the sum comprising only connected diagrams.
The connected correlation functions then admit the $1/N$-expansion,
$$
N^{s-2} \Bigl\langle\prod_{i=1}^s(\tr H^{k_i})\Bigr\rangle^{\mathrm{conn}}
=\sum_{g=0}^\infty N^{-2g} \Bigl\langle\prod_{i=1}^s(\tr H^{k_i})\Bigr\rangle_g^{\mathrm{conn}},
$$
to segregate its part 
where
\be
\Bigl\langle\prod_{i=1}^s(\tr H^{k_i})\Bigr\rangle_g^{\mathrm{conn}} = 
 |\widehat\Gamma_g(k_1, \ldots, k_s)^c|, 
\label{sg2}
\ee
is the part corresponding to the set $\widehat\Gamma_g(k_1, \ldots, k_s)^c$ of
connected fat graphs of genus $g$ with ciliated vertices.

For nonciliated vertices,  we then have the following formula
\bea
(-1)^s\left\langle\prod_{i=1}^s\tr\log(1-u_iH)\right\rangle_g^{\mathrm{conn}}&=&\sum_{\{k_1,\dots,k_s\}\in {\mathbb Z}_+^s}
\prod_{i=1}^s\Bigl(\frac{u_i^{k_i}}{k_i}\Bigr)\left\langle\prod_{i=1}^s(\tr H^{k_i})\right\rangle_g^{\mathrm{conn}}\nonumber\\
&=&\sum_{\{k_1,\dots,k_s\}\in {\mathbb Z}_+^s} \sum_{\gamma \in \Gamma_g(k_1,\dots,k_s)^c} \frac {1}{|\hbox{Aut\,}(\gamma) |}
\prod_{i=1}^s u_i^{k_i},
\label{det-rep}
\eea
where $ \Gamma_g(k_1,\dots,k_s)^c$ is the set of connected fat graphs of genus $g$ with $s$ nonciliated ordered vertices of 
valencies $k_1,\ldots,k_s$ and $\hbox{Aut\,}(\gamma)$ is the automorphism group of the fatgraph $\gamma$ with ordered vertices.
We pass from expressions with nonciliated vertices to those with ciliated vertices, or chord diagrams, by differentiation:
\beq
\label{loop-rep}
\left\langle\prod_{i=1}^s\tr\frac{1}{I-u_iH}\right\rangle_g^{\mathrm{conn}}\equiv
\left\langle\prod_{i=1}^s\tr\biggl[\sum_{k_i=1}^\infty u_i^{k_i}H^{k_i}\biggr]\right\rangle_g^{\mathrm{conn}}=
(-1)^s\biggl[\prod_{i=1}^s u_i\frac{\partial}{\partial u_i}\biggr] \left\langle\prod_{i=1}^s\tr\log(1-u_iH)\right\rangle_g^{\mathrm{conn}}.
\eeq
 By combining formula (\ref{sg2}) with (\ref{loop-rep}), we find that 
 $$\biggl[\prod_{i=1}^s u_i\frac{\partial}{\partial u_i}\biggr] \left\langle\prod_{i=1}^s\tr\log(1-u_iH)\right\rangle_g^{\mathrm{conn}} = \sum_{\gamma\in \widehat\Gamma_{g,s}^c} N^{2-2g}\prod_{i=1}^s u_i^{k_i}.$$

\subsection{Summing up planar subgraphs---formulating the matrix model}
\label{ss:summinf-up}

\begin{figure}[h]
{\psset{unit=1}
\begin{pspicture}(-7,-1)(7,1)
\pcline[linewidth=1pt](-6,0)(-5,0)
\rput(-5.5,-0.7){\makebox(0,0){$1$}}
\rput(-4.5,0){\makebox(0,0){$+$}}
\pcline[linewidth=1pt](-4,0)(-2.5,0)
\rput(-3.25,-0.7){\makebox(0,0){$u^2$}}
\rput(-2,0){\makebox(0,0){$+$}}
\psarc[linewidth=1.5pt,linestyle=dashed](-3.25,0){.4}{0}{180}
\pcline[linewidth=1pt](-1.5,0)(0.2,0)
\rput(-.65,-0.7){\makebox(0,0){$u^4$}}
\rput(0.7,0){\makebox(0,0){$+$}}
\psarc[linewidth=1.5pt,linestyle=dashed](-1,0){.3}{0}{180}
\psarc[linewidth=1.5pt,linestyle=dashed](-.3,0){.3}{0}{180}
\pcline[linewidth=1pt](1.2,0)(2.7,0)
\rput(1.95,-0.7){\makebox(0,0){$u^4$}}
\rput(3.2,0){\makebox(0,0){$+$}}
\psarc[linewidth=1.5pt,linestyle=dashed](1.95,0){.3}{0}{180}
\psarc[linewidth=1.5pt,linestyle=dashed](1.95,0){.5}{0}{180}
\pcline[linewidth=1pt](3.5,0)(5.6,0)
\rput(4.5,-0.7){\makebox(0,0){$u^6$}}
\rput(6.5,0){\makebox(0,0){$+\dots\equiv$}}
\psarc[linewidth=1.5pt,linestyle=dashed](4.15,0){.3}{0}{180}
\psarc[linewidth=1.5pt,linestyle=dashed](4.15,0){.45}{0}{180}
\psarc[linewidth=1.5pt,linestyle=dashed](5.05,0){.35}{0}{180}
%
\psframe[linewidth=1pt](7.5,-0.1)(8.6,0.1)
\pcline[linewidth=1pt](7.4,-0.1)(8.7,-0.1)
\pcline[linewidth=1pt](7.4,0.1)(8.7,0.1)
\rput(8,-0.7){\makebox(0,0){$f(u)$}}
\end{pspicture} }
\caption{\small Summing up rainbow diagrams of chords (dashed lines) for a
single backbone (a solid line). The result is the new (thickened) edge of the backbone.}
\label{fi:rainbow}
\end{figure}
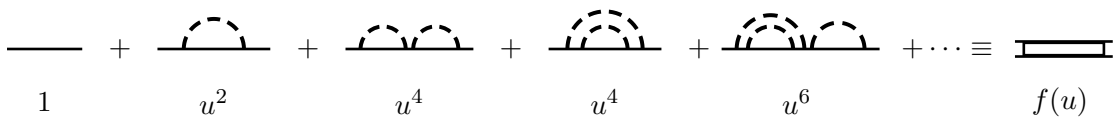

We first perform a partial resummation over planar subgraphs in (\ref{det-rep}). A planar  chord diagram on an interval
is a {\it rainbow diagram} (see examples in Fig.~\ref{fi:rainbow}). Rainbow diagrams with a given number of
chords are enumerated by the Catalan numbers whose generating function is
\beq
\label{factor-f}
f(u_i):=\frac{1-\sqrt{1-4u_i^2}}{2u_i^2},
\eeq
so we effectively replace the original edge of a chord diagram by a thickened edge carrying
the factor $f(u_i)$ thus stripping out all ``pimps,'' or rainbow subgraphs. 

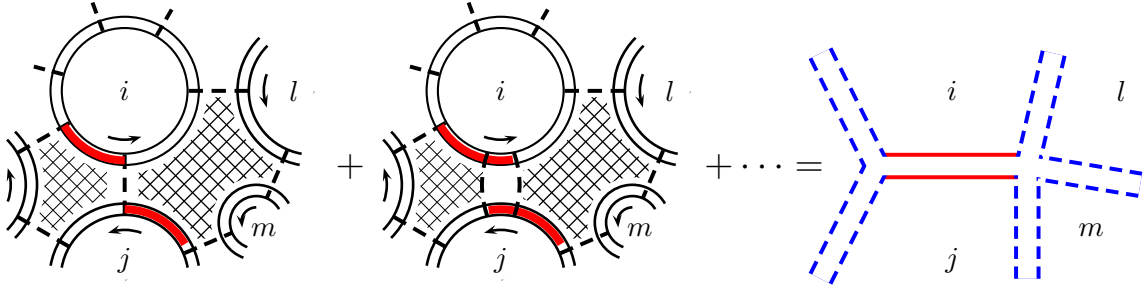
\begin{figure}[h]
{\psset{unit=0.5}
\begin{pspicture}(-10,-5)(4,2.5)
\newcommand{\PATTERN}{%
\psframe[linewidth=1pt,linecolor=white,fillstyle=crosshatch,hatchwidth=0.5pt](0.3,-0.1)(4,-4)
\psframe[linewidth=1pt,linecolor=white,fillstyle=crosshatch,hatchwidth=0.3pt](-0.4,-0.1)(-3,-5)
\pswedge[linecolor=white,fillstyle=solid,fillcolor=white](0,-5){2.2}{0}{180}
\pswedge[linecolor=white,fillstyle=solid,fillcolor=white](5,0){2.2}{180}{270}
\pswedge[linecolor=white,fillstyle=solid,fillcolor=white](0,0){2.2}{180}{360}
\pswedge[linecolor=white,fillstyle=solid,fillcolor=white](3.5,-3.5){1.2}{30}{250}
\pswedge[linecolor=white,fillstyle=solid,fillcolor=white](-4.33,-2.5){2.2}{-45}{45}
\rput{30}(-2.15,-1.25){\psframe[linecolor=white,fillstyle=solid,fillcolor=white](-1,-0.2)(1,2)}
\rput{-30}(-2.15,-3.75){\psframe[linecolor=white,fillstyle=solid,fillcolor=white](-1,0.2)(1,-2)}
\pcline[linewidth=1.5pt,linestyle=dashed](5,0)(3.5,-3.5)
\pcline[linewidth=1.5pt,linestyle=dashed](0,-5)(3.5,-3.5)
\pscircle[linecolor=white,fillstyle=solid,fillcolor=white](3.5,-3.5){.7}
\pswedge[linecolor=white,fillstyle=solid,fillcolor=white](0,-5){1.7}{0}{90}
\pswedge[linecolor=white,fillstyle=solid,fillcolor=white](5,0){1.7}{180}{270}
\pscircle(0,0){2}
\pscircle(0,0){1.7}
\psarc[linewidth=1pt]{->}(0,0){1.3}{250}{290}
\psarc[linewidth=1pt](0,-5){2}{10}{170}
\psarc[linewidth=1pt](0,-5){1.7}{10}{170}
\psarc[linewidth=1pt]{->}(0,-5){1.3}{70}{110}
\psarc[linewidth=1pt](-4.33,-2.5){2}{-45}{45}
\psarc[linewidth=1pt](-4.33,-2.5){1.7}{-45}{45}
\psarc[linewidth=1pt]{->}(-4.33,-2.5){1.3}{-20}{20}
\psarc[linewidth=1pt](5,0){2}{135}{260}
\psarc[linewidth=1pt](5,0){1.7}{135}{260}
\psarc[linewidth=1pt]{->}(5,0){1.3}{160}{200}
\psarc[linewidth=1pt](3.5,-3.5){1}{30}{250}
\psarc[linewidth=1pt](3.5,-3.5){.7}{30}{250}
\psarc[linewidth=1pt]{->}(3.5,-3.5){.4}{90}{200}
\pcline[linewidth=1.5pt,linestyle=dashed](1.7,0)(3.3,0)
\pcline[linewidth=1.5pt,linestyle=dashed](-1.47,-0.85)(-2.85,-1.65)
\pcline[linewidth=1.5pt,linestyle=dashed](-1.47,-4.15)(-2.85,-3.35)
\rput{60}(0,0){\pcline[linewidth=1.5pt,linestyle=dashed](1.7,0)(2.5,0)}
\rput{110}(0,0){\pcline[linewidth=1.5pt,linestyle=dashed](1.7,0)(2.5,0)}
\rput{165}(0,0){\pcline[linewidth=1.5pt,linestyle=dashed](1.7,0)(2.5,0)}
\rput(0,0){\makebox(0,0)[cc]{$i$}}
\rput(0,-4.5){\makebox(0,0)[cc]{$j$}}
\rput(4.5,0){\makebox(0,0)[cc]{$l$}}
\rput(3.7,-3.7){\makebox(0,0)[cc]{$m$}}
}
\rput(-5,0){\PATTERN}
\rput(5,0){\PATTERN}
\rput(1,-2){\makebox(0,0)[cc]{\Large $+$}}
\rput(12,-2){\makebox(0,0)[cc]{\Large $+\cdots=$}}
\rput(-5,0){\pcline[linewidth=1.5pt,linestyle=dashed](0,-1.7)(0,-3.3)}
\rput(-5,0){\psarc[linewidth=3pt,linecolor=red](0,0){1.85}{210}{270}}
\rput(-5,0){\psarc[linewidth=3pt,linecolor=red](0,-5){1.85}{30}{90}}
\rput(5,0){\psarc[linewidth=3pt,linecolor=red](0,0){1.85}{210}{280}}
\rput(5,0){\psarc[linewidth=3pt,linecolor=red](0,-5){1.85}{30}{100}}
\rput(5,0){\psframe[linewidth=1pt,linecolor=white,fillstyle=solid,fillcolor=white](0.05,-2.1)(0.6,-2.9)}
\rput(5,0){\psarc[linewidth=1.5pt,linestyle=dashed](2,-2.5){2.5}{160}{200}}
\rput(5,0){\psarc[linewidth=1.5pt,linestyle=dashed](-2,-2.5){2.5}{-20}{20}}
\rput(17,0){
\pcline[linewidth=10pt,linecolor=red](-2,-2)(2,-2)
\pcline[linewidth=10pt,linecolor=blue,linestyle=dashed](2,-2)(2.7,1)
\pcline[linewidth=10pt,linecolor=blue,linestyle=dashed](2,-2)(2,-5)
\pcline[linewidth=10pt,linecolor=blue,linestyle=dashed](2,-2)(5,-2.5)
\pcline[linewidth=10pt,linecolor=blue,linestyle=dashed](-2,-2)(-3.5,1)
\pcline[linewidth=10pt,linecolor=blue,linestyle=dashed](-2,-2)(-3.5,-5)
\pcline[linewidth=7pt,linecolor=white](-2,-2)(2,-2)
\pcline[linewidth=7pt,linecolor=white](2,-2)(2.7,1)
\pcline[linewidth=7pt,linecolor=white](2,-2)(2,-5)
\pcline[linewidth=7pt,linecolor=white](2,-2)(5,-2.5)
\pcline[linewidth=7pt,linecolor=white](-2,-2)(-3.5,1)
\pcline[linewidth=7pt,linecolor=white](-2,-2)(-3.5,-5)
\rput(0,0){\makebox(0,0)[cc]{$i$}}
\rput(0,-4.5){\makebox(0,0)[cc]{$j$}}
\rput(4.5,0){\makebox(0,0)[cc]{$l$}}
\rput(3.7,-3.7){\makebox(0,0)[cc]{$m$}}
}
\end{pspicture} }
\caption{\small Performing a resummation over ladder diagrams. The thickened edges
associated with the selected ladder, which becomes an edge of a new fat graph,
 are painted dark. The crosshatched domains will become the respective three- and four-valent vertices of the
new fat graph representing a \emph{shape}.}
\label{fi:cycles}
\end{figure}

We next proceed to summing up ladder-type diagrams, where
a ``rung'' of the ladder joins two cycles that carry (either distinct or coinciding) indices
$i$ and $j$ (see an example in Fig.~\ref{fi:cycles}).
Each ladder contains at least one rung, which is
a chord carrying the factor $u_iu_j$. 
We obtain an effective fat graph with new {\em edges} and
{\em vertices} by blowing up cycles of thickened backbone edges
until they will be joined pairwise along rungs (each containing at least one rung);
disjoint parts of these cycles will then constitute loops of lengths $2r_k\ge 6$ alternatively bounded by $r_k$ rungs
(the chords) and
$r_k$ thickened edges of circular backbones;
these loops then become \emph{vertices} of the respective orders $r_k\ge 3$ of the {\em new fat graph}. 

Introducing $e^{\lambda_i}=\frac{1+\sqrt{1-4u_i^2}}{2u_i}$,\quad\hbox{or}\quad $u_i=\frac{1}{e^{\lambda_i}+e^{-\lambda_i}}$,
for each ladder subgraph, we have a sum
\beq
\label{KP-rung}
\sum_{k=1}^\infty (u_iu_jf(u_i)f(u_j))^k=\frac{1}{(u_if(u_i)u_jf(u_j))^{-1}-1}:=\frac{1}{e^{\lambda_i+\lambda_j}-1},
\eeq

We therefore attain the effective description. 

\begin{theorem}\label{lm:diagrams}{\rm\cite{ACNP}}
The genus-$g$ term of the (nonciliated) $s$-backbone connected diagrams is given by the following
(finite!) sum over fatgraph shapes $\Gamma_{g,s}$ of genus $g$ with $s$ faces whose
vertices have valences at least three:
\beq
\Bigl\langle\prod_{i=1}^s\tr\log(e^{\lambda_i}+e^{-\lambda_i}-H)\Bigr\rangle_g^{\mathrm{conn}}
=\sum_{{\mathrm{all\ fatgraphs} }\atop \gamma~\in~\Gamma_{g,s}}
\frac{1}{| \Aut (\gamma)|} \prod_{\mathrm{edges}}\frac{1}{\e^{\lambda_e^{(+)}+\lambda_e^{(-)}}-1}
:= F^{(g)}_s(\lambda_1,\dots,\lambda_s),
\label{new-model}
\eeq
where $\pm$ denotes the two sides (faces) of the edge $e$.
The quantity $F^{(g)}_s(\lambda_1,\dots,\lambda_s)$ in the right-hand side is the term in the
diagrammatic expansion of the free energy of the
Kontsevich--Penner matrix model~\cite{ChMak1} described by the normalised integral
over Hermitian $N\times N$-matrices $X$:
\beq
{\mathcal Z}[\Lambda]:=e^{\sum_{g,s}N^{2-2g}(\alpha/2)^{2-2g-s}F^{(g)}_s(\lambda)}=
\frac{\int DX e^{-\alpha N\tr\bigl[\frac14 \Lambda X\Lambda X+\frac12\log(1-X)+X/2 \bigr]}}
{\int DX e^{-\alpha N\tr\bigl[\frac14 \Lambda X\Lambda X-\frac14 X^2\bigr]}}.
\label{KPMM-1}
\eeq
Here the sum ranges all stable curves $(2g+s>2)$ and $\Lambda=\hbox{diag\,}\bigl(e^{\lambda_1},\dots, e^{\lambda_N}\bigr)$.
\end{theorem}

Differentiating the relation (\ref{new-model}) w.r.t.\ $\lambda_i$ in the right-hand side we obtain the standard
loop means, or (connected) correlation functions $W_s^{(g)}(x_1,\dots,x_s)$, $x_i=e^{\lambda_i}+e^{-\lambda_i}$,
of the Gaussian matrix model enjoying the standard topological recursion relations~\cite{Ey},~\cite{ChEy}. We therefore obtain
the exact relation between resolvents and terms of the expansion of the KPMM free energy:
\beq
W_s^{(g)}(e^{\lambda_1}+e^{-\lambda_1},\dots,e^{\lambda_s}+e^{-\lambda_s})=
\prod_{i=1}^s\left[\frac{1}{e^{\lambda_i}-e^{-\lambda_i}}\frac{\pa}{\pa \lambda_i}\right]F^{(g)}_s(\lambda_1,\dots,\lambda_s).
\label{loopmeans-KPMM}
\eeq
The quantities $W_s^{(g)}(x_1,\dots,x_s)$ here
enjoy the standard topological recursion~\cite{ChEy}, \cite{AlMM} for the spectral curve $x=e^{\lambda}+e^{-\lambda}$,
$y=\frac12 \bigl(e^{\lambda}-e^{-\lambda}\bigr)$.

\section{Kontsevich--Penner matrix model and discrete moduli spaces}
\label{s:KPMM}
\setcounter{equation}{0}

\subsection{The Kontsevich matrix model}\label{ss:Kontsevich}

We turn now to the cell decomposition of moduli spaces of Riemann surfaces of genus $g$ with
$s>0$ marked points proved independently by Harer \cite{HarVir} using Strebel differentials~\cite{Strebel} and by Penner \cite{Penner},~\cite{PenPer} using hyperbolic geometry.  This cell decomposition theorem
states that strata in the cell decomposition of the direct
product ${\mathcal M}_{g,s}\times {\mathbb R}^s_{+}$ of the open moduli space and the $s$-dimensional space of
strictly positive perimeters of holes
are in one-to-one correspondence with fat graphs of genus $g$ with $s$ faces (those are the shapes from
Sec.~\ref{s:matrix-model}) whose edges are decorated with strictly positive numbers $l_i\in {\mathbb R}_{+}$.
The perimeters $P_I$, $I=1,\dots,s$ are the sums of $l_i$ taken (with multiplicities) over edges incident to the
corresponding face (boundary component, or hole).  So it is natural to call them the {\em lengths} of the corresponding edges.

The fundamental theorem by Kontsevich~\cite{Kon88} establishes the relation between the
\emph{intersection indices} $\left\langle\tau_{d_1}\cdots\tau_{d_s}\right\rangle_g;=\int_{\overline{\mathcal M}_{g,s}}\prod_{I=1}^s \psi_I^{d_I}$ and the Kontsevich matrix-model integral. Here $\psi_I$ is a $\psi$-class, or a Chern class, associated with the
$I$th marked point, and integrals of these classes (intersection indices) do not depend on  actual values of $P_I$ being purely cohomological objects.
Multiplying every $\psi_I^{d_I}$ by
$P_I^{2d_I}$ and performing the {\em Laplace transformation} w.r.t. all $P_I$,  we obtain
\beq
\iint_0^\infty dP_1\cdots dP_s e^{-\sum_I P_I\lambda_I}
\int_{\overline{\mathcal M}_{g,s}}\prod_{I=1}^s P_I^{2d_I}\psi_I^{d_I}=
\left\langle\tau_{d_1}\cdots\tau_{d_s}\right\rangle_g \prod_{I=1}^s \prod_{I=1}^s \frac{(2d_I)!}{\lambda_I^{2d_I+1}}.
\label{Kontsevich}
\eeq
Using the explicit representation of $\psi$-classes from~\cite{Kon88} we can present
the left-hand side of (\ref{Kontsevich}) as the sum over three-valent fat graphs with the
weights $1/(\lambda_{I_1}+\lambda_{I_2})$ on edges where $I_1$ and $I_2$ are indices of two (possibly coinciding)
cycles incident to a given edge. Also a factor $2^{|L|-|V|}$ appears (where $|V|$ and $|L|$ are the cardinalities
of the respective sets of vertices and edges).
The generating function is then the celebrated {\em Kontsevich matrix model}
\beq
e^{\sum_{g=0}^\infty\sum_{s=1}^\infty N^{2-2g}\alpha^{2-2g-s} {\mathcal F}^{(g,s)}_{\text{K}}(\{\xi_k\})}:=
\frac{\int DX e^{-\alpha N\tr\bigl[\frac12 X^2\Lambda +X^3/6 \bigr]}}
{\int DX e^{-\alpha N\tr\bigl[\frac12 X^2\Lambda\bigr]}},
\label{Kont}
\eeq
where
\beq
\xi_k:=\frac1N\sum_{i=1}^N\frac{(2k)!}{\lambda_i^{2k+1}} 
=\frac1N\sum_{i=1}^N\int_0^\infty dP_|\,P_I^{2k}e^{-\lambda_IP_I}
\label{times-K}
\eeq
are the {\em times} of the Kontsevich matrix model.

\subsection{Open discrete moduli spaces and KPMM}

As was proposed in~\cite{Ch93}, we set all the lengths of edges of the Penner--Strebel graphs to be nonnegative {\em integers}
$l_i\in {\mathbb Z}_+$, $i=1,\dots, |L|\le 6g-6+3s$. Instead of integrations over ${\mathcal M}_{g,s}$ we take
summations over integer points inside ${\mathcal M}_{g,s}$.

Because the length $l_i$ of every edge appears exactly twice in the sum $\sum_{I=1}^s P_I$,
this sum is always a positive even number, and we must take this restriction into account
when performing the discrete Laplace transformations with the measure $e^{-\sum_{I=1}^s \lambda_I P_I}$. By analogy with 
the continuous Laplace transformation in the Kontsevich model, we introduce 
the new times
\beq
T^\pm_{2k}(\lambda_I):=\frac{\pa^{2k}}{\pa \lambda^{2k}_I}\frac{1}{\mp e^{\lambda_I} -1}=\sum_{P_I=1}^\infty (\mp 1)^{P_I} P_I^{2k} 
e^{-\lambda_I P_I}
\label{times-KP}
\eeq
as discrete Laplace transforms; the above ${\mathbb Z}_2$ restrictions ensure the existence of two sets of times.

Following \cite{Norbury} we thus define the {\em \dvol\ } 
$N_{g,s}(P_1,\dots,P_s)$ which is a weighted count of the integer points inside
${\mathcal M}^{\disc}_{g,s}\times {\mathbb Z}_+^{s}$ for fixed positive integers $P_I$, $I=1,\dots,s$, which are
the perimeters of the holes (cycles). These \dvol\ are equal (modulo the standard factors of volumes of automorphism
groups) to the numbers of all fat graphs with vertices of valencies three and higher and with positive integer
lengths of edges subject to the restriction that the lengths of all cycles (the perimeters) are fixed. Using the identity
$\sum_{I=1}^s \lambda_I P_I=\sum_{e\in L} l_e(\lambda_{I^{(e)}_1}+\lambda_{I^{(e)}_2}),$
where $l_e$ is the length of the $e$th edge and $I^{(e)}_1$ and $I^{(e)}_2$ are the indices of two
(possibly coinciding) cycles incident to the $e$th edge, we obtain that
\beq
\sum_{\{P_I\}\in {\mathbb Z}_+^{s}}N_{g,s}(P_1,\dots,P_s)e^{-\sum_{I=1}^s P_I\lambda_I}
=\sum_{\Gamma_{g,s}}\frac{1}{|\hbox{Aut\,}\Gamma_{g,s}|}\prod_{e=1}^{|L|}
\frac{1}{e^{\lambda_{I^{(e)}_1}+\lambda_{I^{(e)}_2}}-1}.
\label{gen-fun-Norbury}
\eeq
We recognize in (\ref{gen-fun-Norbury}) the genus expansion of the KPMM (\ref{KPMM-1}).  We thus have the lemma
\begin{lemma}\label{lm:Norbury1}{\rm \cite{ACNP}}
The generating function for the Laplace transformed \dvol\  $N_{g,s}(P_1,\dots,P_s)$ is the
KPMM (\ref{KPMM-1}). The correspondence (\ref{gen-fun-Norbury})  is given by the formula
\beq
e^{\sum'_{g,s,P_j\in {\mathbb Z}_+}N^{2-2g}\alpha^{2-2g-s} N_{g,s}(P_1,\dots,P_s)e^{-\sum_{I=1}^s P_I\lambda_I}}
=\frac{\int DX e^{-\alpha N\tr\bigl[\frac12 \Lambda X\Lambda X+\log(1-X)+X \bigr]}}
{\int DX e^{-\alpha N\tr\bigl[\frac12 \Lambda X\Lambda X-\frac12 X^2\bigr]}},
\label{KPMM-Catalan}
\eeq
where the sum ranges all stable curves with $2g-2+s>0$ and strictly positive perimeters $P_l$.
\end{lemma}

\begin{remark}\label{rm:Mulase}
The formula (\ref{KPMM-Catalan}) is valid at all values of $N$ and $\lambda_l$. Specializing it to the case $N=1$ (when
we have just an ordinary integral instead of the matrix one) and setting $\lambda_l=\lambda$, $\alpha=1/\hbar$, and $x=e^{\lambda}+e^{-\lambda}$, we obtain
$$
e^{\sum'_{g,s,P_j\in {\mathbb Z}_+}\hbar^{2g+s-2} N_{g,s}(P_1^2,\dots,P_s^2)e^{-\lambda\sum_{I=1}^s P_I}}
=\sqrt{\frac{1-e^{-2\lambda}}{\pi\hbar}}e^{ -({2\hbar})^{-1}e^{2\lambda}+\hbar^{-1}\lambda}F(\hbar,x),
$$
where the function
$$
F(\hbar,x):= \int_{-\infty}^\infty dt\, e^{-\frac{1}{\hbar}(t^2/2+xt+\log t)}
$$
satisfies the second-order differential equation
$$
\Bigl[\hbar^2 \frac{\partial^2 }{\partial x^2}+x \hbar\frac{\partial }{\partial x}+(1-\hbar)\Bigr] F(\hbar,x)=0.
$$
We thus reproduce the equation of the quantum curve from \cite{DM14}.
\end{remark}

Note that the \dvol\  are quasi-polynomials: their coefficients depend on the mutual parities of the
$P_I$'s and we present one more proof of this fact below (see Corollary~\ref{cor:polynom}).
Because the generating function (\ref{KPMM-1})
is related by (\ref{loopmeans-KPMM}) to the standard $s$-loop Gaussian means $W_s^{(g)}$, we have the following lemma.

\begin{lemma}\label{lm:Norbury}{\rm\cite{ACNP}.}
The correlation functions $W_s^{(g)}(x_1,\dots,x_s)$ of the Gaussian matrix model subject to the
standard topological recursion based on
the spectral curve $x=e^{\lambda}+e^{-\lambda}$, $y=\frac{1}{2}(e^{\lambda}-e^{-\lambda})$ are related to
the \dvol\  by the following explicit relation:
\beq
W_s^{(g)}(e^{\lambda_1}+e^{-\lambda_1},\dots,e^{\lambda_s}+e^{-\lambda_s})
=\prod_{I=1}^s \left[\frac{1}{e^{\lambda_I}-e^{-\lambda_I}}
\sum_{P_I=1}^\infty P_Ie^{-P_I\lambda_I}\right] N_{g,s}(P_1,\dots,P_s).
\label{Norbury-rel}
\eeq
\end{lemma}

The matrix model (\ref{KPMM-1}) manifests many remarkable
properties. Besides being the generating function for the \dvol\  related to Gaussian means, it is also
equivalent~\cite{ChMak2,KMMMZ} to the Hermitian matrix model with the potential determined by the Miwa change of the variables 
$t_k=\frac{1}{k}\tr (e^\Lambda+e^{-\Lambda})^{-k}+\frac12 \delta_{k,2}$, it is the generating function for the number of clean Belyi functions, or for 
the corresponding Grothendieck {\em dessins d'enfant} \cite{AmCh14} (see also \cite{AMMN})
and, finally, in the special times $T^{\pm}_{2r}$, $r=0,1,\dots$, (\ref{times-KP}), it is equal to the product of two Kontsevich matrix
models \cite{Ch95}, intertwined by a canonical transformation of the variables. We now turn to this last property.

\begin{lemma}\label{lm:canonical}{\rm(\cite{Ch95})}
The partition function ${\mathcal Z}[\Lambda]$ (\ref{KPMM-1})
expressed in the times $T^\pm_{k}(\lambda)$ (\ref{times-KP}) depends only on the even times $T^\pm_{2k}(\lambda)$
and satisfies the following exact relation:
\beq
{\mathcal Z}[\Lambda]=e^{{\mathcal F}_{\text{KP}}[\{T^\pm_{2n}\}]}
=e^{C(\alpha N)}e^{-N^{-2}{\mathcal A}}e^{{\mathcal F}_{\text{K}}[\{T^+_{2n}\}]+{\mathcal F}_{\text{K}}[\{T^-_{2n}\}]},
\label{canonical}
\eeq
where ${\mathcal F}_{\text{K}}[\{T^\pm_{2n}\}]$ is a free energy of the
Kontsevich matrix model (\ref{Kont}), $T^\pm_{2n}$ given by (\ref{times-KP}) are therefore the times of the KdV hierarchies, 
and ${\mathcal A}$ is the canonical transformation operator
\bea
{\mathcal A}&=&\sum_{m,n=0}^\infty \frac{B_{2(n+m+1)}}{4(n+m+1)}\frac{1}{(2n+1)!(2m+1)!}
\Bigl\{\frac{\pa^2}{\pa T^+_{2n}\pa T^+_{2m}}+\frac{\pa^2}{\pa T^-_{2n}\pa T^-_{2m}}+
2(2^{2(n+m+1)}-1)\frac{\pa^2}{\pa T^+_{2n}\pa T^-_{2m}}\Bigr\}\nonumber\\
&{}&+\sum_{n=2}^\infty \alpha N^2\frac{2^{2n-1}}{(2n+1)!}\Bigl(\frac{\pa}{\pa T^-_{2n}}+\frac{\pa}{\pa T^+_{2n}} \Bigr).
\label{A}
\eea
Here $C(\alpha N)$ is a function depending only on $\alpha N$ that ensures that
${\mathcal F}_{\text{KP}}[\{T^\pm_{2n}\}]=0$ for $T^\pm_{2n}\equiv 0$ and $B_{2k}$ are the Bernoulli numbers
generated by $t/(e^t-1)=\sum_{m=0}^\infty B_mt^m/(m!)$.
\end{lemma}

From this canonical transformation we immediately obtain the (ordinary) graph representation for the
term ${\mathcal F}_{g,s}[\{T^\pm_{2n}\}]$ of the expansion of
$$
{\mathcal F}_{\text{KP}}[\{T^\pm_{2n}\}]=\sum_{g,s} N^{2-2g}\alpha^{2-2g-s} {\mathcal F}_{g,s}[\{T^\pm_{2n}\}].
$$

\begin{lemma}\label{lm:graph}{\rm(\cite{Ch95,ACNP})}
We can present a term ${\mathcal F}_{g,s}[\{T^\pm_{2n}\}]$ of the genus expansion of the KPMM (\ref{KPMM-1}) as
a sum of a finite set of graphs $G_{g,s}$ described below; each graph contributes the factor also described below divided 
by the order of the automorphism group of the graph. 
\begin{itemize}
\item each node (a vertex) $v_i$, $i=1,\dots,q$, of
a graph $G_{g,s}$ is decorated by the marking "$+$" or "$-$", by the genus $g_i\ge 0$, and has $s_i$
endpoints of edges
incident to it ($2g_i-2+s_i>0$, i.e., all nodes are stable);
each endpoint of an edge carries a nonnegative integer $k^{\pm}_{r,i}$, $r=1,\dots,s_i$; these
integers are subject to restriction that $\sum_{r=1}^{s_i}k^{\pm}_{r,i}=3g_i-3+s_i$
where the superscript $+$ or $-$ is determined by the marking of the vertex;
\item edges can be external legs (ordinary leaves) with $k^{\pm}_{r,i}\ge 0$ (we let $a_i\ge 0$ denote the number
of such legs incident to the $i$th vertex),
half-edges (dilaton leaves) with $k^{\pm}_{r,i}\ge 2$ (we let $b_i\ge 0$ denote the number
of such legs incident to the $i$th node), or internal edges incident either to two different nodes or to the same node (their two endpoints carry in general different numbers $k^{\pm}_{r_1,i_1}$ and $k^{\pm}_{r_2,i_2}$) (we let $l_i$ denote the number of internal edge endpoints
incident to the $i$th node);
\item each node contributes the Kontsevich intersection index
$
\Bigl\langle \tau_{k^{\pm}_{1,i}}\cdots \tau_{k^{\pm}_{s_i,i}}\Bigr\rangle_{g_i};
$
\item every internal edge with endpoint markings $(k_1^{+},k_2^{+})$ or $(k_1^{-},k_2^{-})$
(two endpoints of such an edge can be incident to the same node)
contribute the factor
$$
-\frac{B_{2(k^{\pm}_1+k^{\pm}_2+1)}}{2(k^{\pm}_1+k^{\pm}_2+1)}
\frac{1}{(2k^{\pm}_1+1)!(2k^{\pm}_2+1)!}
$$
and every internal edge with endpoint markings $(k_1^{+},k_2^{-})$
(two endpoints of such an edge can be incident only to distinct nodes having different markings $+$ and $-$)
contributes the factor
$$
-\frac{B_{2(k^+_1+k^-_2+1)}}{2(k^+_1+k^-_2+1)}
\frac{2^{2(k^+_1+k^-_2+1)}-1}{(2k^+_1+1)!(2k^-_2+1)!};
$$
\item every half-edge with the marking $r^{\pm}\ge 2$ contributes the factor $-\frac{2^{2r^{\pm}-1}}{(2r^{\pm}+1)!}$;
\item every external leg with the marking $k^{\pm}_{r,i}$ contributes the corresponding time
$T^{\pm}_{2k^{\pm}_{r,i}}$;
\item $\sum_{i=1}^q (g_i+l_i/2-1)+1=g$ (the total genus $g$ is equal to the sum of internal genera plus the
number of loops in the graph);
\item $\sum_{i=1}^q a_i=s$ (the total number of external legs is fixed and equal to $s$);
\end{itemize}
From the above formulas, we have that
\beq
\sum_{j=1}^s k_j^{\text{Ext}}=3g-3+s-\sum_{j=1}^{|L|}(1+k^{\text{Int}}_{j,1}++k^{\text{Int}}_{j,2})
-\sum_{j=1}^{|B|}(k^{\text{Half}}_j-1),
\label{reducing}
\eeq
where, disregarding the node labels,
$k_j^{\text{Ext}}\ge 0$ are indices of the external edges, $k^{\text{Int}}_{j,1}\ge 0$ and $k^{\text{Int}}_{j,2}\ge 0$
are indices of endpoints of the internal edges, $k^{\text{Half}}_j\ge 2$ are indices of half-edges, and $|L|$ and
$|B|$ are the cardinalities of the respective sets of internal edges and half-edges of the graph.
\end{lemma}

\begin{figure}[h]
{\psset{unit=0.7}
\begin{pspicture}(-7,-3)(5,3)
\newcommand{\PATTERN}{%
{\psset{unit=1}
\pscircle[fillstyle=solid,fillcolor=yellow](0,0){1}
\pscircle[fillstyle=solid,fillcolor=white,linecolor=white](0,0){0.4}
\psline[linewidth=2pt,linecolor=blue]{->}(-0.8,0)(-1.5,0)
}
}
\rput(-2,2){\PATTERN}
\rput(-3.6,2){\makebox(0,0)[rc]{\small $T^+_{6g_1-4}$}}
\rput(-2,-2){\PATTERN}
\rput(-3.6,-2){\makebox(0,0)[rc]{\small $T^-_{6g_2-2}$}}
\psline[linewidth=2pt,linecolor=blue]{-|}(-2,2.8)(-2,3.3)
\rput(-2,2.8){\makebox(0,0)[rt]{$\small 3^+$}}
\pscircle[fillstyle=solid,fillcolor=yellow](1,0){1}
\pscircle[fillstyle=solid,fillcolor=white,linecolor=white](1,0){0.4}
\rput(1.55,0.6){\makebox(0,0)[ct]{\small $1^+$}}
\rput(1.55,-0.6){\makebox(0,0)[cb]{\small $0^+$}}
\rput(-2,2){\makebox(0,0)[cc]{\small $g_1^+$}}
\rput(-2,-2){\makebox(0,0)[cc]{\small $g_2^-$}}
\rput(1,0){\makebox(0,0)[cc]{\small $0^+$}}
\psarc[linecolor=red,linestyle=dashed,linewidth=2pt](2,0){0.8}{-120}{120}
\psline[linecolor=red,linestyle=dashed,linewidth=2pt](-2,1.2)(-2,-1.2)
\rput(-2,1.2){\makebox(0,0)[rb]{\small $0^+$}}
\rput(-2,-1.2){\makebox(0,0)[rt]{\small $0^-$}}
\psline[linecolor=red,linestyle=dashed,linewidth=2pt](-1.25,1.5)(0.25,0.5)
\rput(-1.35,1.6){\makebox(0,0)[cb]{\small $0^+$}}
\rput(0.35,0.5){\makebox(0,0)[lc]{\small $0^+$}}
\psline[linecolor=red,linestyle=dashed,linewidth=2pt](-1.25,-1.5)(0.25,-0.5)
\rput(-1.35,-1.6){\makebox(0,0)[ct]{\small $1^-$}}
\rput(0.35,-0.5){\makebox(0,0)[lc]{\small $0^+$}}
\end{pspicture} }
\caption{\small The typical diagram from the graph expansion $G_{g,s}$.}
\label{fi:primer}
\end{figure}
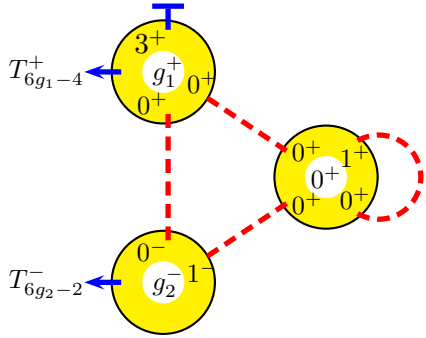

The proof is just another application of Wick's theorem, now in the form of exponential of a 
linear-quadratic differential operator (\ref{A}); for the typical form in the above sum, see Fig.~\ref{fi:primer}.

This lemma immediately implies the corollary
\begin{corollary}\label{cor:polynom}
The quantities ${\mathcal F}_{g,s}[\{T^\pm_{2n}\}]$ are polynomials such that, for every monomial
$T^+_{2n_1}\cdots T^-_{2n_s}$ we have that $\sum_{i=1}^s n_i\le 3g-3+s$, and the highest term
with $\sum_{i=1}^s n_i= 3g-3+s$ is
$$
\bigl\<\tau_{n_1}\cdots \tau_{n_s}\bigr\>_g\Bigl(\prod_{i=1}^s T^+_{2n_i}+\prod_{i=1}^s T^-_{2n_i}\Bigr).
$$
This also implies that all \dvol\  $N_{g,s}(P_1,\dots,P_s)$ are ${\mathbb Z}_2$-quasi-polynomials in $P^2_I$.
\end{corollary}

\noindent
{\bf Proof.}  The \dvol\  $N_{g,s}(P_1,\dots,P_s)$
depend only on even powers of $P_I$ because ${\mathcal F}_{g,s}$ depend only on even times
$T^\pm_{2n}$; the quasi-polynomiality follows immediately from the fact that ${\mathcal F}_{g,s}$ are
polynomials in $T^+_{2n}$ and $T^-_{2n}$.

\begin{remark}\label{rm:alternative}
Note that the quadratic part of the differential operator (\ref{A}) manifests the alternating structure 
because the Bernoulli numbers $B_{2n}$ are positive for odd $n$ and negative for even $n$,
$$
B_{2n}=(-1)^{n+1} \frac{2(2n)!}{(2\pi)^{2n}}\Bigl[1+\frac{1}{2^{2n}}+\frac{1}{3^{2n}}+\cdots\Bigr].
$$
\end{remark}

\subsection{The times for the multi-resolvents}\label{ss:multi-times}

We first consider $N_{g,1}(P)$, which are polynomials of degree $3g-2$ in $P^2$, are nonzero only for even $P$, and must vanish for all $P=2,\dots,4g-2$ (because the minimum number of edges of the genus $g$ shape with one face is $2g$, and
the minimum nonzero $P$ is therefore $4g$).
We thus have that, for even $P$, $N_{g,1}(P)$ has the form $\prod_{k=1}^{2g-1}\bigr(P^2-(2k)^2\bigl)\mathop{Pol}_{g-1}(P^2)$,
where $\mathop{Pol}_{g-1}(x)$ is a polynomial of degree $g-1$ and $N_{g,1}(P)$ vanishes for odd $P$, so
its Laplace transform in formula (\ref{Norbury-rel}) is
\beq
W_1^{(g)}(e^{\lambda}+e^{-\lambda})=\frac{-1}{e^{\lambda}-e^{-\lambda}}
\sum_{i=0}^{g-1}\frac{b_{i}^{(g)}}{2^{4g+2i-1}(4g+2i-1)!}
\prod_{k=1}^{2g+i-1}\left(\frac{\pa^{2}}{\pa \lambda^{2}}-(2k)^2\right)\frac{\pa}{\pa \lambda}\frac{1}{e^{2\lambda}-1}
\label{W1-1}
\eeq
for some coefficients $b_{i}^{(g)}$. 
Using that $-\frac{\pa}{\pa \lambda}\frac{1}{e^{2\lambda}-1}
=\frac{2}{(e^{\lambda}-e^{-\lambda})^2}$
and the relation
\beq
\left(\frac{\pa^{2}}{\pa \lambda^{2}}-(m)^2\right)\frac{1}{(e^{\lambda}-e^{-\lambda})^{m}}
=\frac{4(m)(m+1)}{(e^{\lambda}-e^{-\lambda})^{m+2}},\quad m\ge1,
\label{recursion1}
\eeq
we obtain the general representation for the one-loop mean,
\be
W_1^{(g)}(e^{\lambda}+e^{-\lambda})=\frac{1}{e^{\lambda}-e^{-\lambda}}
\sum_{i=0}^{g-1}b_{i}^{(g)}
\frac{1}{(e^{\lambda}-e^{-\lambda})^{4g+2i}}=\frac{1}{(e^{\lambda}-e^{-\lambda})^{4g+1}}
\sum_{i=0}^{g-1}\frac{b_{i}^{(g)}}{(e^{\lambda}-e^{-\lambda})^{2i}}.
\label{W1g-B}
\ee
In \cite{ACNP}, we have found the transition formulas between $b_{i}^{(g)}$ and the coefficients $P_{g,i}$ from
\cite{ACRPS}: the integrality of $b_{i}^{(g)}$ implies that of $P_{g,i}$ and vice versa, but the positivity conjecture for
$P_{g,s}$ put forward in \cite{ACRPS} requires an additional work.

We now consider the general $s$-resolvent case.
From (\ref{loopmeans-KPMM}) we have that
the (stable) loop means (with $2g+s-2\ge1$) are polynomials 
$W_s^{(g)}(e^{\lambda_1}+e^{-\lambda_1},\dots,e^{\lambda_s}+e^{-\lambda_s})
=F_{g,s}\bigl(\{t^{\pm}_{2n_j+1}(\lambda_{j})\}\bigr)$ in times obtained by the substitution 
\beq
\label{times}
T^{\pm}_{2d}\to t^{\pm}_{2d+1}(\lambda_{j}):=
\frac{1}{e^{\lambda_j}-e^{-\lambda_j}}
\left(\frac{\partial}{\partial \lambda_j}\right)^{2d+1}\frac{1}{e^{\lambda_i}\pm 1},
\eeq
All the times $t^{\pm}_{2d+1}(\lambda)$ are strictly
skew-symmetric with respect to the change of variables $\lambda\to -\lambda$.

Using (\ref{W1-1}) and the fact that
\beq
\label{time-w}
t^{-}_{2d+1}(\lambda)+t^{+}_{2d+1}(\lambda)=
\frac{1}{e^{\lambda}-e^{-\lambda}}
\left(\frac{\partial}{\partial \lambda}\right)^{2d+1}\frac{2}{e^{2\lambda}-1}
=\sum_{j=1}^{d+1}q_{j,d}\frac{1}{(e^\lambda-e^{-\lambda})^{2j+1}}
\eeq
and
\beq
\label{time-as}
t^{-}_{2d+1}(\lambda)-t^{+}_{2d+1}(\lambda)=
\frac{1}{e^{\lambda}-e^{-\lambda}}
\left(\frac{\partial}{\partial \lambda}\right)^{2d+1}\frac{2}{e^{\lambda}-e^{-\lambda}}
=\sum_{j=1}^{d+1}\tilde q_{j,d}\frac{e^\lambda+e^{-\lambda}}{(e^\lambda-e^{-\lambda})^{2j+1}}
\eeq
with some integer coefficients $q_{j,d}$ and $\tilde q_{j,d}$, where relation (\ref{time-as}) follows from
that $\frac{1}{e^\lambda-1}+\frac{1}{e^\lambda+1}=\frac{2}{e^\lambda-e^{-\lambda}}$
and from another useful representation
\bea
\frac{1}{e^{\lambda}-e^{-\lambda}}\frac{\partial}{\partial\lambda}
\prod_{k=1}^{d}\left(\frac{\pa^{2}}{\pa \lambda^{2}}-(2k-1)^2\right)\frac{2}{e^\lambda-e^{-\lambda}}
&=&\frac{1}{e^{\lambda}-e^{-\lambda}}\frac{\partial}{\partial\lambda}
\frac{2^{2d+1} (2d)!}{(e^\lambda-e^{-\lambda})^{2d+1}}\nonumber\\
&=&-2^{2d+1}(2d+1)!
\frac{e^\lambda+e^{-\lambda}}{(e^\lambda-e^{-\lambda})^{2d+3}},
\label{W1-as}
\eea
we can equivalently expand $F_{g,s}\bigl(\{t^{\pm}_{2n_j+1}(\lambda_{j})\}\bigr)$
in  the variables
\beq
\label{sj}
s_{k,\beta}(\lambda):=\frac{(e^{\lambda}+e^{-\lambda})^\beta_{}}{{(e^{\lambda}-e^{-\lambda})^{2k+3}}},
\quad k=0,\dots,3g+s-3, \ \beta=0,1.
\eeq
In the next section we demonstrate that the coefficients of these expansions
are related to the ancestor invariants of a CohFT.

We now present the general structure of the multiloop means.
\begin{lemma}\label{lm:multiloop}
The general expression for a stable ($2g+s-3\ge 0$) loop mean
$W_s^{(g)}(e^{\lambda_1}+e^{-\lambda_1},\dots,e^{\lambda_s}+e^{-\lambda_s})$ in terms of
the variables $s_{k,\beta}(\lambda)$ given by (\ref{sj}) reads:
\beq
W_s^{(g)}(e^{\lambda_1}+e^{-\lambda_1},\dots,e^{\lambda_s}+e^{-\lambda_s})
=\sum_{\vec{k},\vec{\beta}}\widehat{b}^{(g)}_{\vec{k},\vec{\beta}}\prod_{j=1}^s s_{k_j,\beta_j}(\lambda_j),
\label{W->sk}
\eeq
where $k_j$ and $\beta_j$ are subject to the restrictions:
\beq
2g-1+\frac12\sum_{j=1}^s\beta_j\le \sum_{j=1}^s k_j\le 3g+s-3,
\qquad \sum_{j=1}^s\beta_j=0\ \hbox{mod}\ 2.
\label{restr-k}
\eeq
The two nonstable loop means are
\bea
&{}&W_1^{(0)}(e^{\lambda}+e^{-\lambda})=e^{-\lambda},
\label{W10}\\
&{}&W_2^{(0)}(e^{\lambda_1}+e^{-\lambda_1},e^{\lambda_2}+e^{-\lambda_2})=\prod_{i=1,2}\prod_{j=1,2}
\frac{1}{e^{\lambda_i}-e^{-\lambda_j}}
\label{W20}
\eea
\end{lemma}

We prove restrictions (\ref{restr-k}) using two considerations: first, if we scale $\lambda_j\to \infty$ uniformly for
all $j$, $\lambda_j\to \lambda_j+R$, every edge contributes a factor $e^{-2R}$ plus $s$ factors
$e^{-R}$ due to the derivatives. The minimum number of edges (for a shape with one vertex) is $2g+s-1$, so the minimum
factor appearing is $e^{(-4g-3s+2)R}$ whereas $s_{k,\beta}(\lambda)$ scale as $e^{(-3-2k+\beta)R}$, which results
in the lower estimate. The upper estimate emerges out of the pole behaviour at $\lambda_j = 0$. On the one hand, $s_{k,\beta}(\lambda)\sim \lambda^{-2k-3}$ as $\lambda\to 0$ irrespectively on $\beta$; on the other hand,
from the relation to the Kontsevich model we can conclude that the pole structure of the derivatives of the Kontsevich KdV times is $t_{d_j}(\lambda_j)\sim \lambda_j^{-2d_j-3}$ with $\sum_j d_j\le 3g+s-3$ and therefore
$\sum_j d_j=\sum_j k_j$, which leads to the upper estimate. That the sum of the $\beta_j$ factors is even follows from
the symmetricity of the total expression with respect to the total change of the times $T^{\pm}\to T^{\mp}$; under this change,
the variables $s_{k,\beta}(\lambda)$ behave as $s_{k,\beta}(\lambda)\to (-1)^\beta s_{k,\beta}(\lambda)$,
so the sum of the beta factors must be even.

In Sec.~\ref{s:TR}, we use the topological recursion to prove that all admissible by (\ref{restr-k}) coefficients
$\widehat b^{(g)}_{\vec k,\vec \beta}$ are positive integers (see Theorem~\ref{thm-integral}).

\section{Cohomological field theory from \dvol}\label{s:CohFT}
\setcounter{equation}{0}

We now describe a cohomological field theory (CohFT) associated to the \dvol.  A dimension $d$ Frobenius manifold structure is equivalent to a CohFT  for a dimension $d$ vector space $H$ with a basis $\{e_{\alpha}\}$ and a metric $\eta$.  We show that the quasi-polynomial \dvol\  are equivalent to the correlation functions of the CohFT associated to the Hurwitz Frobenius manifold $H_{0,(1,1)}$ described in the introduction.  We give two accountings of the genus 0 case: the first approach is constructive and the other generalises to all genera.  The constructive approach also implies that we deal with a homogenous CohFT.  The primary correlation functions of our CohFT turn out to be virtual Euler characteristics 
$\chi(\modm_{g,n})$ of moduli spaces.

\subsection{Cohomological field theories}\label{ss:Frob}

Given a complex vector space $H$ equipped with a complex metric $\eta$, a CohFT is a sequence of $S_s$-equivariant linear maps
$$
I_{g,s}:H^{\otimes s}\to H^*(\overline{\modm}_{g,s}),
$$
which satisfy the following compatibility conditions with respect to inclusion of strata.
Any partition into two disjoint subsets $I\sqcup J=\{1,\dots,s\}$ defines a map
$\phi_I:\overline{\modm}_{g_1,|I|+1}\times\overline{\modm}_{g_2,|J|+1}\to\overline{\modm}_{g,s}$ such that
$$
\phi_I^*I_{g,s}(v_1\otimes\cdots\otimes v_s)=I_{g_1,|I|+1}\otimes I_{g_2,|J|+1}
\Bigl(\bigotimes_{i\in I}v_i\otimes\Delta\otimes\bigotimes_{j\in J}v_j\Bigr)
$$
where $\Delta=\sum_{\alpha,\beta}\eta^{\alpha\beta}e_{\alpha}\otimes e_{\beta}$ with respect to a basis $\{e_{\alpha}\}$ of $H$.  
The map $\psi:\overline{\modm}_{g-1,s+2}\to\overline{\modm}_{g,s}$ induces
$$
\psi^*I_{g,s}(v_1\otimes\cdots\otimes v_s)=I_{g-1,s+2}(v_1\otimes\cdots\otimes v_s\otimes\Delta).
$$
The three-point function $I_{0,3}$ together with the metric $\eta$ induces a product $\bullet$ on $H$,
$u\bullet v=\sum_{\alpha,\beta}I_{0,3}(u\otimes v\otimes e_{\alpha})\eta^{\alpha\beta}e_{\beta}$,
where $I_{0,3}$ takes its values in $\bc$.   A vector $e_0$ satisfying
$$
I_{0,3}(v_1\otimes v_2\otimes e_0)=\eta(v_1\otimes v_2),\quad \forall v_1,v_2\in H
$$
is the identity element for the product on $H$.

An extra condition satisfied both by the CohFT under consideration and by Gromov--Witten invariants
pertains to the forgetful map  for $s\geq 3$, $\pi:\overline{\modm}_{g,s+1}\to\overline{\modm}_{g,s}$, which
induces
\be
I_{g,s+1}(v_1\otimes\cdots\otimes v_s\otimes e_0)=\pi^*I_{g,s}(v_1\otimes\cdots\otimes v_s).
\label{forget}
\ee

\subsection{Quasipolynomials and ancestor invariants}\label{ss:quasi}
The \dvol\  $N_{g,s}(P_1,\dots,P_s)$ are mod 2 even quasi-polynomials, i.e. it is an even polynomial on each coset of $2\bz^s\subset\bz^s$.   Define a basis of mod 2 even quasi-polynomials induced (via tensor product) from the following single-variable basis $p_{k,\alpha}(b)$ for $k=0,1,2,\dots$ and $\alpha=0,1$.
$$p_{0,0}(b)=\left\{\begin{array}{cc}1,&b \text{\ even}\\0,&b \text{\ odd}\end{array}\right.,\quad p_{0,1}(b)=\left\{\begin{array}{cc}0,&b \text{\ even}\\1,&b \text{\ odd}\end{array}\right.,\qquad p_{k+1,\alpha}(b)=\sum_{m=0}^b m p_{k,\alpha}(m),\quad k\geq 0.$$
Then
\be
p_{k,\alpha}(b)=\frac{p_{0,k+\alpha}(b)}{4^kk!}\hspace{-.5cm}\prod_{{0<m\le k\atop m=k+\alpha\ (\mathrm{mod\,}2)}}(b^2-m^2)
\label{pk1}
\ee
where in the second subscript we mean $k+\alpha$ (mod 2).

Put $\vec{k}=(k_1,\dots,k_s)$ and  $\vec{\alpha}=(\alpha_1,\dots,\alpha_s)$.
\begin{theorem} \label{th:cohft} We have that
$$N_{g,s}(P_1,\dots,P_s)=\sum_{\vec{k},\vec{\alpha}} c^g_{\vec{k},\vec{\alpha}} \prod_{i=1}^s p_{k_i,\alpha_i}(P_i)$$
where the coefficients are ancestor invariants:
\be
c^g_{\vec{k},\vec{\alpha}}=\int_{\overline{\modm}_{g,s}}I_{g,s}(e_{\alpha'_1}\otimes\cdots\otimes e_{\alpha'_s})\prod_{i=1}^s\psi_i^{k_i}.
\label{ancestor}
\ee 
\end{theorem}
The \emph{proof} is an application of \cite{DOSSIde} where theories with spectral curves satisfying special conditions were identified
with semisimple CohFTs. The outcome of applying \cite{DOSSIde} is non-constructive so we prove the genus zero case in a different way that provides an explicit realisation of the CohFT.

\subsection{A homogeneous CohFT in genus zero}

The {\em primary} correlators of a CohFT are 
$Y_{g,s}:=\int_{\overline{\modm}_{g,s}}I_{g,s}:H^{\otimes s}\to\bc$, and we assemble them into the generating function
$$
F(t_0,...,t_{D-1})=\sum N^{2-2g}\frac{1}{s!}Y_{g,s}=\sum N^{2-2g}F_g
$$
where $(t_0,\dots,t_{D-1})$ in $ H^*$ is the dual basis of $\{e_0,...,e_{D-1}\}$.  The genus 0 part $F_0$ is the 
\emph{prepotential} of the CohFT.


\begin{theorem}[Manin \cite{ManFro} Theorem III.4.3]   \label{th:manin}
One can uniquely reconstruct a genus 0 CohFT from abstract correlation functions.  
\end{theorem}
The Deligne--Mumford compactification $\overline{\modm}_{g,s}$ possesses a natural stratification indexed by {\em dual graphs}. The dual graph of $\Sigma\in\overline{\modm}_{g,s}$ has vertices corresponding to the irreducible components of $\Sigma$ with specified genera, edges corresponding to the nodes (cusps) of $\Sigma$, and a {\em tail}---an edge with an open end (no vertex)---corresponding to each labeled point of $\Sigma$.  If $\Gamma$ is a dual graph of type $(g,s)$, then the collection of curves $D_{\Gamma}$ whose associated dual graph is $\Gamma$ forms a stratum of $\overline{\modm}_{g,s}$.  The closure $\overline{D}_{\Gamma}=\cup_{\Gamma'<\Gamma}D_{\Gamma'}$, where the partial ordering is given by edge contraction, represents an element of $H^*(\overline{\modm}_{g,s})$.   Keel \cite{KeeInt} proved that $H^*(\overline{\modm}_{0,s})$ is generated by $\overline{D}_{\Gamma}$ and derived all relations.

The proof of Theorem~\ref{th:manin} uses that
$$ 
\int_{\overline{D}_{\Gamma}}I_{0,s}(v_1\otimes\cdots\otimes v_s)=\bigotimes_{v\in V_{\Gamma}}Y_{0,|v|}\left(\bigotimes _{i=1}^s v_i\otimes\Delta^{\otimes |E_{\Gamma}|}\right).
$$
which defines evaluation of a cohomology class on boundary strata tautologically from the definition of a CohFT.  Because $H^*(\overline{\modm}_{0,s})$ is generated by its boundary strata, and relations in $H^*(\overline{\modm}_{0,s})$ agree with the relations satisfied by abstract correlation functions, this suffices for proving the theorem.

In particular, we have the primary invariants 
\bea
&{}&Y_{0,3}(e_0\otimes e_0\otimes e_1)=1=Y_{0,3}(e_1\otimes e_1\otimes e_1), \quad Y_{0,s}(e_0\otimes \hbox{anything})=0,\quad 
s>3\nonumber\\
&{}&Y_{0,s}(e_1^{\otimes s})=N_{0,s}(0,...,0)=\chi(\modm_{0,s})\quad s>3
\eea
that define a genus 0 CohFT.  

A CohFT is conformal if its prepotential is quasihomogeneous with respect to the {\em Euler vector field}:
\begin{equation}  \label{euler}
 E\cdot F_0=(3-d)F_0+Q(t)
 \end{equation}
where $Q$ is a quadratic polynomial in $t=(t_0,\dots,t_{D-1})$.  Using the genus 0 reconstruction in Theorem~\ref{th:manin}, Manin proved that a conformal CohFT induces the following push-forward condition on the genus 0 CohFT.

Let $\xi$ be any vector field on $H$ treated as a manifold with coordinates $t_0,\dots,t_{D-1}\in H^*$.   The Lie derivative with respect to $\xi$ of the CohFT correlation functions $I_{g,s}$  induces a natural action
$$
(\xi\cdot I)_{g,s}(v_1\otimes\cdots\otimes v_s)=\deg I_{g,s}(v_1\otimes\cdots\otimes v_s)-\sum_{j=1}^s I_{g,s}(v_1\otimes\cdots \otimes[\xi,v_j]\otimes\cdots \otimes v_s)+\pi_*I_{g,s+1}(v_1\otimes\cdots \otimes v_s\otimes\xi)
$$
where $\pi:\overline{\modm}_{g,s+1}\to\overline{\modm}_{g,s}$ is the forgetful map,
 $I_{g,s}$ are  ($H^*(\overline{\modm}_{0,s})$-valued) tensors on $H$, and the vector field $\xi$ acts infinitesimally on $I_{g,s}$.

A CohFT is {\em homogeneous} of weight $d$ if
\begin{equation}  \label{homog}
(E\cdot I)_{g,s}=((g-1)d+s)I_{g,s}
 \end{equation}
If a preprotential satisfies the homogeneity condition (\ref{euler}), the proof of Theorem~\ref{th:manin} implies that the corresponding
genus 0 CohFT is homogeneous.  The Lie derivative of the bivector $\Delta$ dual to the metric $\eta$ on $H$ can be calculated in flat coordinates
$$
\cl_E\cdot\Delta=\cl_E\cdot\eta^{ij}e_i\otimes e_j=\eta^{ij}([E,e_i]\otimes e_j+e_i\otimes [E,e_j])=(d-2)\eta^{ij}e_i\otimes e_j=(d-2)\Delta
$$
where we have used a choice of flat coordinates \cite{DubGeo} with respect to which $\eta=\delta_{i,D-1-i}$ and
$E=\sum_i(\alpha_it_i+\beta_i)\frac{\partial}{\partial t_i}$, where $\alpha_i+\alpha_{D-1-i}=2-d$.

\subsection{Proof of Theorem~\ref{th:cohft} in genus 0.}

We can now prove the genus 0 case of Theorem~\ref{th:cohft}.  For this we produce a 
prepotential from the primary (constant) terms of $N_{0,s}(P_1,\dots,P_s)$, which uniquely (and constructively) determines a genus 0 CohFT.  Moreover, the quasihomogeneity of the prepotential implies a homogeneous CohFT.  The higher coefficients of $N_{0,s}(P_1,\dots,P_s)$ satisfy a homogeneity condition that makes them the correlation functions of the homogeneous CohFT.

The prepotential
\begin{equation}  \label{prep}
F_0=\sum\frac{1}{s!}Y_{0,s}=\frac{1}{2}t_0^2t_1+\sum_{s\geq 3}\frac{1}{s!}N_{0,s}(\vec{0})t_1^s
=\frac{1}{2}t_0^2t_1+\frac{1}{2}(1+t_1)^2\log(1+t_1)-\frac{1}{2}t_1-\frac{3}{4}t_1^2
\end{equation}
assembled from $N_{0,s}(\vec{0})=(-1)^{s-3}(s-3)!$ is quasihomogeneous with respect to the Euler vector field $E=t_0\frac{\partial}{\partial t_0}+2(1+t_1)\frac{\partial}{\partial t_1}$:
$$E\cdot F_0=4F_0+t_1^2+t_0^2.
$$
This ensures that the genus 0 CohFT $I_{0,s}$ produced from Theorem~\ref{th:manin} satisfies
\begin{equation}  \label{pushf}
\pi_*I_{g,s+1}(e_S\otimes e_1)=\frac{1}{2}\Bigl(1-g+s-\deg-\sum\alpha_{i_k}\Bigr)I_{g,s}(e_S)
\end{equation}
where $e_S=e_{i_1}\otimes...\otimes e_{i_s}$, and $\alpha_0=1$, $\alpha_1=2$ are the coefficients of $E$.  
The CohFT also satisfies the pull-back condition (\ref{forget}).


\begin{theorem}{\rm[Teleman \cite{TelStr}]}
A semi-simple homogenous CohFT with flat identity is uniquely and explicitly reconstructible from genus zero data. 
\end{theorem}
Thus, given the genus 0 primary invariants $N_{0,s}(\vec{0})$ there is a unique homogenous CohFT with flat identity.  Below we demonstrate that its correlation functions agree with the coefficients of $N_{g,s}(P_1,\dots,P_s)$.

The pushforward relation (\ref{pushf}) expressed in terms of correlators is \cite{ACNP}
$$
\int_{\overline{\modm}_{g,s+1}}\hspace{-3mm}I_{g,s+1}(e_S\otimes e_1)\prod_{i=1}^s\psi_i^{k_i}
=\Bigl(\sum_{i=1}^s \frac{k_i}{2}+\chi_{g,s}\Bigr)
\int_{\overline{\modm}_{g,s}}I_{g,s}(e_S)\prod_{i=1}^s\psi_i^{k_i}+\sum_{j=1}^s\int_{\overline{\modm}_{g,s}}I_{g,s}(e_{S\backslash\{j\}}\otimes e^*_j)\prod_{i=1}^s\psi_i^{k_i-\delta_{ij}}.
$$

The condition $E\cdot F_0=4F_0+t_1^2+t_0^2$ on $N_{0,s}(\vec{0})$ is a specialisation to $g=0$ and $P_i=0$ of the divisor equation \cite{NorStr}
\begin{equation}   \label{divisor}
N_{g,s+1}(0,P_1,\dots,P_s)=\sum_{j=1}^s\sum_{k=1}^{P_j-1}kN_{g,s}(P_1,...,P_s)|_{P_j=k}+\left(\frac{1}{2}\sum_{j=1}^s P_j+\chi_{g,s}\right)N_{g,s}(P_1,\dots,P_s).
\end{equation} 

The flat identity pull-back condition is known as the {\em string equation} on correlators for $2g-2+s>0$:
$$\int_{\overline{\modm}_{g,s+1}}I_{g,s+1}(v_1\otimes\cdots\otimes v_s\otimes e_0)\prod_{i=1}^s\psi_i^{k_i}
=\sum_{j=1}^s\int_{\overline{\modm}_{g,s}}I_{g,s}(v_1\otimes\cdots\otimes v_s)\prod_{i=1}^s\psi_i^{k_i-\delta_{i,j}}
$$
and agrees with the recursion \cite{NorStr}
\begin{equation}   \label{string}
N_{g,s+1}(1,P_1,\dots,P_s)=\sum_{j=1}^s\sum_{k=1}^{P_j}kN_{g,s}(P_1,\dots,P_s)|_{P_j=k}
\end{equation}
In particular, this proves the genus 0 case of Theorem~\ref{th:cohft} since the recursions (\ref{divisor}) and (\ref{string}) uniquely determine the correlation functions of $I_{0,s}$ and $N_{0,s}(P_1,\dots,P_s)$.


This constructive proof describes explicitly the genus 0 classes $I_{0,s}(e_S)\in H^*(\overline{\modm}_{0,s})$:
$$
\int_{\overline{\modm}_{0,s}} I_{0,s}(e_S)=\left\{\begin{array}{cc}\chi(\modm_{0,s})&e_S=e_1^{\otimes s}\\ 0&\text{otherwise}\end{array}\right..
$$

\subsection{General proof of Theorem~\ref{th:cohft} using DOSS method \cite{DOSSIde}.}

We establish the correspondence between correlation functions of the CohFT and discrete volumes in higher genera
applying the results of \cite{DOSSIde}, where it was shown that for spectral curves satisfying a compatibility condition, the Givental reconstruction of higher genus correlation functions can be formulated in terms of graphs, and the same graphs can be used to calculate topological recursion.  

Dunin-Barkowsky, Orantin, Shadrin, and Spitz \cite{DOSSIde} using Eynard's technique of \cite{Ey11}
associated to any semi-simple CohFT a local spectral curve $(\Sigma,B,x,y)$.  The Givental
$R$-matrix gives rise to the bidifferential $B$ on the spectral curve 
\be\label{Givental1}
\sum_{p,q}\check{B}^{i,j}_{p,q}z^pw^q=\frac{\delta^{ij}-\sum_{k=1}^NR^i_k(-z)R^j_k(-w)}{z+w}
\ee
where $\check{B}^{i,j}_{p,q}$ are coefficients of an asymptotic expansion of the Laplace transform of the regular part of 
the Bergmann bidifferential $B$ expressed in terms of the local coordinates $s_i=\sqrt{x-x(a_i)}$ where $dx(a_i)=0$.  The $R$-matrix together with the transition matrix $\Psi$ from a flat to a normalised canonical bases expresses the meromorphic differential $ydx$ in terms of $s_i$.  In particular, this implies a compatibility condition (\ref{compatible}) between the differential $ydx$ and the bifferential $B$.  


One can apply \cite{DOSSIde} in either direction, beginning with a semi-simple CohFT or a spectral curve.  The prepotential $F_0$ (\ref{prep}) gives rise to a semi-simple CohFT thus generating the $R$-matrix and the transition matrix $\Psi$ and hence the spectral curve.  But having in hands a candidate for the spectral curve, we can start with the spectral curve and apply \cite{DOSSIde} to obtain the coefficients of $N_{g,s}(P_1,\dots,P_s)$ as ancestor invariants of a CohFT.  Because it agrees with the above CohFT in genus 0, by uniqueness it is the same CohFT produced by Teleman's theorem.

The spectral curves for the \dvol\  and Gromov--Witten invariants of $\bp^1$ are similar:
\be
\begin{array}{llll}
\hbox{\dvol\ } & x=z+1/z, & y=z, & B=\frac{dzdz'}{(z-z')^2}\cr
\hbox{GW invariants} & x=z+1/z, & y=\ln{z}, & B=\frac{dzdz'}{(z-z')^2}
\end{array}
\label{vol-GW}
\ee
and because $x$ and $B$ determine the $R$-matrix uniquely, it is the same for the both curves.  The $R$-matrix for the 
Gromov--Witten invariants of $\bp^1$ reads \cite{DOSSIde}:
$$
R(u)=\sum_{k=0}^{\infty}R_ku^k,\quad R_k=\frac{(2k-1)!!(2k-3)!!}{2^{4k}k!}
\left(\begin{array}{cc}-1&(-1)^{k+1}2ki\\2ki&(-1)^{k+1}\end{array}\right).
$$
The results of \cite{DOSSIde} can be applied to those spectral curves for which a Laplace transform of $ydx$ is related to this $R$-matrix (which is essentially the Laplace transform of the regular part of the bidifferential).

For local coordinates $s_i$, $i=1,2$ near $x=\pm 2$ given by $x=s_i^2\pm 2$ 
$$ y= 1+s_1+\frac{1}{2}s_1^2+\sum_{k=1}^{\infty}(-1)^{k-1}\frac{(2k-3)!!}{2^{3k}k!}s_1^{2k+1},\qquad
y= -1+is_2+\frac{1}{2}s_2^2-i\sum_{k=1}^{\infty}\frac{(2k-3)!!}{2^{3k}k!}s_2^{2k+1},$$
so we obtain
\bea
&{}&
\check{(ydx)}_1
=\frac{\sqrt{u}}{2\sqrt{\pi}}\int_{\gamma_1} e^{-u(x-2)}ydx\sim\sum_{k=0}^{\infty}(-1)^{k-1}\frac{(2k+1)!!(2k-3)!!}{2^{4k+1}k!}u^{-(k+1)}
\nonumber\\
&{}&\check{(ydx)}_2
=\frac{\sqrt{u}}{2\sqrt{\pi}}\int_{\gamma_2} e^{-u(x+2)}ydx\sim-i\sum_{k=0}^{\infty}\frac{(2k+1)!!(2k-3)!!}{2^{4k+1}k!}u^{-(k+1)},
\nonumber
\eea
where $(-1)!!=1$, $(-3)!!=-1$, and we let $\sim$ denote the Poincare asymptotic in the parameter $u$.

The compatibility condition between the differential $ydx$ and the bifferential $B$ reads
\begin{equation}  \label{compatible}
\frac{1}{\sqrt{2}}\left(\begin{array}{cc}1&i\end{array}\right)\cdot\frac{1}{\sqrt{2}}R(u)=\left(\begin{array}{cc}\check{(ydx)}_1&\check{(ydx)}_2\end{array}\right)
\end{equation}
which uses the first row of the transition matrix $\Psi=\frac{1}{\sqrt{2}}\left(\begin{array}{cc}1&i\\1&-i\end{array}\right)$.
A direct verification indicates that it is satisfied for $x=z+1/z$, $y=z$, $B=dzdz'/(z-z')^2$.

From this, \cite{DOSSIde} supplies the times
\be
\xi^0_0=\frac{1}{2}\left(\frac{1}{1-z}-\frac{1}{1+z}\right),\qquad
\xi^1_0=\frac{1}{2}\left(\frac{1}{1-z}+\frac{1}{1+z}\right),\qquad
\xi^i_k=\left(\frac{d}{dx}\right)^k\xi^i_0=\sum_k p_{k,i}z^k
\label{question}
\ee
and the main result
$$W^{(g)}_s(x_1,\dots,x_s)=\sum_{\vec{k},\vec{\alpha}} c^g_{\vec{k},\vec{\alpha}} \prod_{i=1}^n \xi_{k_i,\alpha_i}$$
where the coefficients are the ancestor invariants (\ref{ancestor}).
As remarked above, the CohFT produced this way necessarily coincides with the homogeneous CohFT produced by Teleman's theorem since they both use Givental reconstruction and the same initial data.

\subsection{Ancestor invariants and Gaussian means}\label{ss:Ancestor-Gauss}

Lemma~\ref{lm:Norbury} and formulas (\ref{W1-1}) and (\ref{W1-as}) straightforwardly express
the loop means in terms of  the ancestor invariants.

\begin{theorem}\label{th:Wgs2ckl}
We have the following explicit relation between the ancestor invariants (\ref{ancestor}) of a CohFT
and the Gaussian means:
\beq
W^{(g)}_s\bigl(e^{\lambda_1}+e^{-\lambda_1},\dots,e^{\lambda_s}+e^{-\lambda_s}\bigr)
=\sum_{\vec{k},\vec{\alpha}}c^g_{\vec{k},\vec{\alpha}}\prod_{j=1}^s \widehat p_{k_j,\alpha_j}(\lambda_j),
\eeq
where
\beq
\widehat p_{k,\alpha}(\lambda)=\left\{\begin{array}{lll}
2^{1-2r}(2r+1)s_{r,0}(\lambda),&k=2r,&\alpha=0;\cr
2^{-2r}(2r+1)s_{r,1}(\lambda),&k=2r,&\alpha=1;\cr
2^{-2r+2}2r(2r+1)s_{r,1}(\lambda),&k=2r-1,&\alpha=0;\cr
2^{-2r-1}s_{r,0}(\lambda),&k=2r+1,&\alpha=1,
\end{array}
\right.
\label{pk2}
\eeq
and $s_{r,\beta}(\lambda)$, $\beta=0,1$, are defined in (\ref{sj}).
\end{theorem}

\begin{example}\label{ex:top}
The topological (degree zero) part of the CohFT is
$$I_{g,s}(e_{\alpha_1}\otimes\cdots\otimes e_{\alpha_s})=\epsilon(\vec{\alpha})2^g+\ \hbox{higher degree terms}$$
where $\displaystyle \epsilon(\vec{\alpha})\equiv\sum_{i=1}^s\alpha_i$ (mod 2) is 0 or 1.
This explains the asymptotic behaviour of the topological invariants $W^{(g)}_{s}$ at their poles.
\end{example}
\begin{example}\label{ex:volume}
If $\{e_0,e_1\}$ is a basis of $H$ corresponding to flat coordinates then
$$\int_{\overline{\modm}_{g,s}}I_{g,s}(e_1^{\otimes s})=\chi(\modm_{g,s}).$$
This uses the fact that $N_{g,s}(0,0,\dots,0)=\chi(\modm_{g,s})$ and
$$p_{k,\alpha}(0)=\left\{\begin{array}{cc} 1,&(k,\alpha)=(0,0)\\
 0,& \text{otherwise}.\end{array}\right.
 $$
\end{example}

{\sl We thus identify the coefficients $\widehat{b}^g_{\vec{k},\vec{\beta}}$ of the expansions (\ref{W->sk}) with (linear 
combinations) of the ancestor invariants $c^g_{\vec{k},\vec{\alpha}}$ using the identification} (\ref{pk2}): for $s=1$, we have
\bea
&{}&\widehat b^g_{r,0}=2^{1-2r}(2r+1)c^g_{2r,0}+2^{-1-2r}c^g_{2r+1,1},\nonumber\\
&{}&\widehat b^g_{r,1}=2^{-2r}(2r+1)c^g_{2r,1}+2^{2-2r}2r(2r+1)c^g_{2r-1,0},\nonumber
\eea
and for general $s$ we have up to $2^s$ terms $c^{(g)}_{\vec k,\vec \alpha}$ with all admissible substitutions
$(k_i,1)\leftrightarrow(k_i-1,0)$.

\section{The topological recursion}\label{s:TR}
\setcounter{equation}{0}

\def\Col#1{
\begin{pspicture}(0.2,1)
\multiput(0,0)(0,0.2){#1}{\psframe(0,0)(0.2,0.2)}
\end{pspicture}
}

\def\ColOne#1{
\begin{pspicture}(0.2,1)
\multiput(0,0)(0,0.2){#1}{\psframe(0,0)(0.2,0.2)}
\rput(0,-0.2){\makebox(0,0)[lb]{\tiny$1$}}
\end{pspicture}
}

\def\ColP#1{
\begin{pspicture}(0.2,1)
\multiput(0,0)(0,0.2){#1}{\psframe(0,0)(0.2,0.2)}
\rput(0,-0.2){\makebox(0,0)[lb]{\tiny $p$}}
\end{pspicture}
}

\def\Colg#1{
\begin{pspicture}(0.2,1)
\multiput(0,0)(0,0.2){#1}{\psframe[fillstyle=solid,fillcolor=lightgray](0,0)(0.2,0.2)}
\end{pspicture}
}

\def\ColgOne#1{
\begin{pspicture}(0.2,1)
\multiput(0,0)(0,0.2){#1}{\psframe[fillstyle=solid,fillcolor=lightgray](0,0)(0.2,0.2)}
\rput(0,-0.2){\makebox(0,0)[lb]{\tiny$1$}}
\end{pspicture}
}

\def\ColgP#1{
\begin{pspicture}(0.2,1)
\multiput(0,0)(0,0.2){#1}{\psframe[fillstyle=solid,fillcolor=lightgray](0,0)(0.2,0.2)}
\rput(0,-0.2){\makebox(0,0)[lb]{\tiny $p$}}
\end{pspicture}
}

\def\Cols#1{
\begin{pspicture}(0.2,0.6)
\multiput(0,0)(0,0.2){#1}{\psframe(0,0)(0.2,0.2)}
\end{pspicture}
}

\def\ColsOne#1{
\begin{pspicture}(0.2,0.6)
\multiput(0,0)(0,0.2){#1}{\psframe(0,0)(0.2,0.2)}
\rput(0,-0.2){\makebox(0,0)[lb]{\tiny$1$}}
\end{pspicture}
}

\def\ColsP#1{
\begin{pspicture}(0.2,0.6)
\multiput(0,0)(0,0.2){#1}{\psframe(0,0)(0.2,0.2)}
\rput(0,-0.2){\makebox(0,0)[lb]{\tiny $p$}}
\end{pspicture}
}

\def\Colsg#1{
\begin{pspicture}(0.2,0.6)
\multiput(0,0)(0,0.2){#1}{\psframe[fillstyle=solid,fillcolor=lightgray](0,0)(0.2,0.2)}
\end{pspicture}
}

\def\ColsgOne#1{
\begin{pspicture}(0.2,0.6)
\multiput(0,0)(0,0.2){#1}{\psframe[fillstyle=solid,fillcolor=lightgray](0,0)(0.2,0.2)}
\rput(0,-0.2){\makebox(0,0)[lb]{\tiny$1$}}
\end{pspicture}
}

\def\ColsgP#1{
\begin{pspicture}(0.2,0.6)
\multiput(0,0)(0,0.2){#1}{\psframe[fillstyle=solid,fillcolor=lightgray](0,0)(0.2,0.2)}
\rput(0,-0.2){\makebox(0,0)[lb]{\tiny $p$}}
\end{pspicture}
}

\begin{figure}[h]
{\psset{unit=1}
\begin{pspicture}(-6,-2)(6,2)
\setlength{\fboxsep}{5pt}
\setlength{\fboxrule}{2pt}
\newcommand{\PATTERN}{%
{\psset{unit=1}
\pscircle*(0,0){0.1}
\pscircle[linewidth=1pt,linestyle=dashed](0,0){0.3}
\psline[linewidth=2pt,linecolor=blue]{->}(-0.8,0)(-0.1,0)
\rput(-0.7,0.1){\makebox(0,0)[rb]{$p_1$}}
\rput(0,0.4){\makebox(0,0)[cb]{$q$}}
}
}
\rput(-5,0){
\psline[linewidth=2pt,linecolor=blue](-0.5,0.3)(0,0.3)
\psline[linewidth=2pt,linecolor=blue](-0.5,0.1)(0,0.1)
\psline[linewidth=2pt,linecolor=blue](-0.5,-0.1)(0,-0.1)
\psline[linewidth=2pt,linecolor=blue](-0.5,-0.3)(0,-0.3)
}
\rput(-5,0){\makebox(0,0)[lc]{\framebox{$W^{(g)}_s$}\ =\ $\sum\limits_{k=2}^{s}$}}
\rput(-5.6,0.4){\makebox(0,0)[rc]{$p_1$}}
\rput(-5.6,0.1){\makebox(0,0)[rc]{$\vdots$}}
\rput(-5.6,-0.4){\makebox(0,0)[rc]{$p_s$}}
\rput(0,0.4){
\rput(-1,0){\psline[linewidth=2pt,linecolor=blue](-0.8,.8)(0,0)
\rput(-0.8,0.9){\makebox(0,0)[cb]{$p_k,\overline p_k$}}
\psline[linewidth=2pt,linecolor=blue](-0.7,-0.4)(0.8,-0.4)
\psline[linewidth=2pt,linecolor=blue](-0.7,-1.2)(0.8,-0.8)
\psline[linewidth=2pt,linecolor=blue](0,0)(0.8,0)
\psline[linewidth=2pt,linecolor=blue](-0.8,.8)(0,0)
\PATTERN
}
\rput(-0.25,-0.4){\makebox(0,0)[lc]{\framebox{$W^{(g)}_{s-1}$}}}
\rput(-1.8,-0.4){\makebox(0,0)[rc]{$p_2$}}
\rput(-1.8,-0.7){\makebox(0,0)[rc]{$\vdots$}}
\rput(-1.8,-1.2){\makebox(0,0)[rc]{$p_s$}}
\rput(-2,-0.8){\makebox(0,0)[rc]{$\widehat p_k$}}
\rput(0,-1.5){\makebox(0,0)[cc]{CP}}
}
\rput(1.3,0){\makebox(0,0)[lc]{$+$}}
\rput(0,0.3){
\rput(3,0){\psarc[linewidth=2pt,linecolor=blue](0.8,-1.36){1.6}{90}{120}
\psarc[linewidth=2pt,linecolor=blue](0.8,1.36){1.6}{240}{270}
\psline[linewidth=2pt,linecolor=blue](-0.7,-0.4)(0.8,-0.4)
\psline[linewidth=2pt,linecolor=blue](-0.7,-0.8)(0.8,-0.6)
\psline[linewidth=2pt,linecolor=blue](-0.7,-1.2)(0.8,-0.8)
\PATTERN}
\rput(4,0){
\rput(-0.25,-0.3){\makebox(0,0)[lc]{\framebox{$W^{(g-1)}_{s+1}\biggm.$}}}
\rput(-1.8,-0.4){\makebox(0,0)[rc]{$p_2$}}
\rput(-1.8,-0.7){\makebox(0,0)[rc]{$\vdots$}}
\rput(-1.8,-1.2){\makebox(0,0)[rc]{$p_s$}}
\rput(0,-1.5){\makebox(0,0)[cc]{PI}}
}}
\rput(6,0){\makebox(0,0)[lc]{$+\hskip-8pt \sum\limits^{\phantom{X}}_{{I\sqcup J\atop =\{p_2,\dots,p_s\}}}$}}
\rput(0.7,0){
\rput(8,0){\psarc[linewidth=2pt,linecolor=blue](0.8,-1.36){1.6}{90}{120}
\psarc[linewidth=2pt,linecolor=blue](0.8,1.36){1.6}{240}{270}
\psline[linewidth=2pt,linecolor=blue](-0.7,-0.4)(0.8,-0.4)
\psline[linewidth=2pt,linecolor=blue](-0.7,-0.8)(0.8,-0.8)
\psline[linewidth=2pt,linecolor=blue](-0.7,0.8)(0.8,0.8)
\PATTERN}
\rput(9,0){
\rput(-0.25,0.6){\makebox(0,0)[lc]{\framebox{$W^{(g_1)}_{s_1+1}$}}}
\rput(-0.25,-0.6){\makebox(0,0)[lc]{\framebox{$W^{(g_2)}_{s_2+1}$}}}
\rput(-1.8,0.8){\makebox(0,0)[rc]{$I$}}
\rput(-1.8,-0.7){\makebox(0,0)[rc]{$J$}}
\rput(0,-1.5){\makebox(0,0)[cc]{PII}}
}}
\end{pspicture} }
\caption{\small The graphical representation of the recursion in (\ref{recursion}). It clearly resembles a \emph{coproduct} (the first term in the right-hand side)
and \emph{product} (the second and third terms in the right-hand side) operations; arrowed line depicts the recursion kernel $K(p_1,q)$ and
nonarrowed lines depicts Bergmann kernels $B(p_k,q)$ and $B(\overline p_k,q)$.}
\label{fi:recursion}
\end{figure}
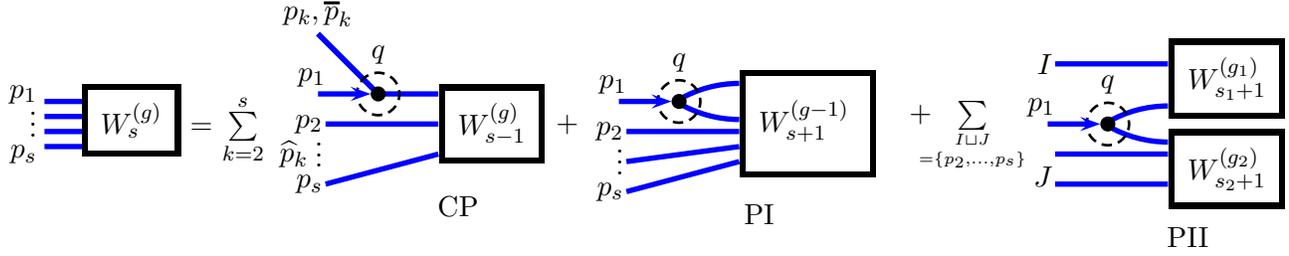

In this section, we present the main ingredients of the topological recursion method developed in \cite{Ey,ChEy,CEO,EOrInv}. In parallel, we adapt the
general construction to the Gaussian means $W^{(g)}_s(x_1,\dots,x_s)$:
\begin{itemize}
\item[{\bf(i)}] The input is a spectral curve $\Sigma_{x,y}=0$ with two meromorphic differentials, $dx$ and $dy$, on this curve. The zeros of $dx$ are the
\emph{branching points}. For the Gaussian means, this curve is the sphere $yx-y^2=1$, and we use the convenient 
local coordinates:
\be
x=e^{\lambda}+e^{-\lambda}, \qquad y=e^{\lambda},\qquad dx=(e^{\lambda}-e^{-\lambda})d\lambda.
\label{XY}
\ee
In the Gaussian mean case, we 
consider the covering of this sphere by two maps: $y=e^{\lambda}$ and $\overline y=e^{-\lambda}$; the sphere is represented as a cylinder
obtained from the strip $\Im \lambda\in [0,2\pi]$ by identifying points $(x,0)$ of the real line $\Im\lambda=0$
with the points $(x,2i\pi)$ of the line $\Im\lambda=2\pi$. We have two branching points
$\lambda=0,i\pi$.
\item[{\bf(ii)}]  The next ingredient is the Bergmann 2-differential $B(p,q)$ that is a symmetric differential with zero $A$-cycles (which are absent in a genus zero case under consideration here) and
with double poles along the diagonal $p=q$. We also need its antiderivative $E(p,q)$ which is a 1-differential in $p$ and a function of $q$ defined 
as $\int_{\overline q}^q B(p,\bullet)$. For the Gaussian means, 
\be
B(p,q)=\frac{d e^\lambda d e^{\mu}}{(e^\lambda-e^\mu)^2}, \qquad E(p,q)=\frac{d e^\lambda}{e^\lambda-e^\mu}, \quad p=e^\lambda,\ 
q=e^\mu.
\label{Bpq}
\ee
\item[{\bf(iii)}]
We define the \emph{recursion kernel} $K(p,q)$ to be the $(1,-1)$-differential $K(p,q)=E(p,q)\frac{1}{(y(q)-\overline y(q))dx}$; 
for the Gaussian means,
\be
K(p,q)=\frac{d e^\lambda}{e^\lambda-e^\mu}\frac{1}{(e^\mu-e^{-\mu})^2 d\mu}, \quad p=e^\lambda,\ 
q=e^\mu,
\label{Kpq}
\ee
where one of the factors $e^\mu-e^{-\mu}$ in the denominator comes from the difference $y(q)-\overline y(q)$ and the another comes
from $dx$. 
\item[{\bf(iv)}]
We introduce the \emph{correlation functions} $W^{(g)}_s(p_1,\dots p_s)$ to be symmetric $s$-differentials determined recurrently as follows:
We choose one of the variables, $p_1$, as a \emph{root}. Then,
\be
W^{(0)}_3(p_1,p_2,p_3)=\sum_{\mathrm{res\,}dx=0} K(p_1,q)[B(p_2,q)+B(\overline p_2,q)][B(p_3,q)+B(\overline p_3,q)],
\label{W03}
\ee
\be
W^{(1)}_1(p_1)=\sum_{\mathrm{res\,}dx=0} K(p_1,q)B(q,\overline q),
\label{W11}
\ee
\bea
&{}&W^{(g)}_s(p_1,p_2,\dots, p_s)=\sum_{\mathrm{res\,}dx=0} K(p_1,q)
\biggl[ \sum_{k=2}^s\bigl[B(p_k,q)+B(\overline p_k,q)\bigr]W^{(g)}_{s-1}(q,p_2,\dots, \widehat p_k,\dots, p_s)\biggr.\nonumber\\
&{}&\qquad\quad\biggl.
+W^{(g-1)}_{s+1}(q,q,p_2,\dots,p_s)+\hskip-0.3cm
\sum_{{g_1+g_2=g\atop I\sqcup J=\{p_2,\dots, p_s\}}}\hskip-0.8cm \biggm.^\prime \hskip 0.4cm
W^{(g_1)}_{|I|+1}(q,\{p_i\}_{i\in I}) W^{(g_2)}_{|J|+1}(q,\{p_j\}_{j\in J})\biggr],
\label{recursion}
\eea
where the right-hand side is explicitly symmetric w.r.t. $p_2,\dots, p_s$ but not w.r.t. $p_1$, $\sum'$ means that we take only stable terms (those
with $2g-2+s>0$) explicitly segregating the only nonstable contribution  (the term with $[B(p_k,q)+B(\overline p_k,q)\bigr]$). The hat over a
symbol indicates its omission from the list of arguments and in the last term we take the sum over all partitions of the set of arguments
$\{p_2,\dots,p_s\}$ into two nonintersecting subsets $I$ and $J$. We depict the recursion relation schematically in Fig.~\ref{fi:recursion}.
\end{itemize}

Using (\ref{recursion}) we construct all higher $W^{(g)}_s$ out of $W^{(0)}_3(p_1,p_2,p_3)$ and $W^{(1)}_1(p_1)$. 

The lemma from \cite{ChEy} states that, although the recursion relation (\ref{recursion}) is not explicitly symmetric w.r.t. permutations of
$p_1,\dots, p_s$, the whole sum in the right-hand side of (\ref{recursion}) is actually symmetric.

\subsection{The topological recursion for the Gaussian means}

In any local theory satisfying the topological recursion, all stable $W^{(g)}_s(x_1,\dots,x_s)$ have singularities only at the branch points.
In the Gaussian case, we therefore conclude that the only singularities in the right-hand side of (\ref{recursion}) besides poles of high orders at
the branching points (for $W^{(g)}_s$, the highest possible order of a pole is $6g+2s-3$)
 are simple poles at $q=p_1$ arising from $K(p_1,q)$ and double poles at $q=p_k$, $q=\overline p_k$
arising from $[B(p_k,q)+B(\overline p_k,q)]$. We can thus perform the integration w.r.t. $q$ in the right-hand side by evaluating residues at these points, not
at the branch points, which drastically simplifies actual calculations. This will allow us to formulate the result in terms of operations on Young diagrams.

\begin{definition}
For each (stable) pair $(g,s)$ we introduce the following set of admissible pairs of Young diagrams $(D_1,D_2)$ (as usual, we let
$l(D)$ and $|D|$ denote the respective length (the number of columns) and volume (the number of boxes) of a Young diagram $D$):
\bea
&{}&Y^{(g)}_s=\{(D_0,D_1)| \ s=l(D_0)+l(D_1),\ l(D_1)\in 2{\mathbb Z}_{\ge0},\nonumber\\
&{}&\phantom{yuyjjuoioioyuyuyu}2g-1+s+l(D_1)/2\le |D_0|+|D_1|\le 3g+2s-3\}.
\label{You1}
\eea

The Young diagram $D_0$ comprises say $d_r$ columns of positive integer heights $t_r$ such that $t_r>t_s$ for $r<s$ and hence $|D_0|=\sum_r d_rt_r$.
Correspondingly, $D_1$ comprises say $k_j$ columns
of heights $r_j$, $r_j>r_i$ for $j<i$, and hence $|D_1|=\sum_j k_jr_j$. 
\end{definition}

When we depict a pair of Young diagrams $(D_0,D_1)$, we color $D_0$ white and $D_1$ grey. If one of the diagrams is empty we simply omit them from the figure.

\begin{definition}\label{def-hom}
We define a homomorphism 
$$ 
{\mathcal F} : \mathbb Z[Y^{(g)}_{s} ] \rightarrow R
$$ 
to the space $R$ of rational functions of $e^{\lambda_i}$, $i=1,\dots,s$ considered as formal complex variables
obtained by mapping each pair $(D_0,D_1)\in Y^{(g)}_{s} $ to the function 
\beq
{\mathcal F}(D_0,D_1):=\sum_{{L_1\sqcup L_2\sqcup\cdots \sqcup R_1\sqcup R_2
\sqcup\cdots=\{1,\dots,s\}\atop |L_i|=d_i,\ |R_j|=k_j}}\prod_{i} \Bigl( \prod_{\alpha\in L_i}
s_{t_{i}-1,0}(\lambda_\alpha)\Bigr) \prod_{j} \Bigl( \prod_{\gamma\in R_j} s_{r_j-1,1}(\lambda_\gamma)\Bigr),\label{hom-F}
\eeq
where the sum ranges over all partitions of the set $\{1,\dots,s\}$ into disjoint subsets $R_i$, $L_j$ with the respective cardinalities 
$d_i$, $k_j$
and $s_{k,\beta}(\lambda):=(e^\lambda+e^{-\lambda})^\beta(e^\lambda-e^{-\lambda})^{-2k-3}$.
\end{definition}

\begin{theorem}\label{thm-Gaussian}
The Gaussian means $W^{(g)}_s(x_1,\dots, x_s)$ are the coefficients of  the differentials
$$
W^{(g)}_s(p_1,p_2,\dots, p_s)=W^{(g)}_s(x_1,\dots, x_s)dx_1\cdots dx_s,
$$
where
$$
W^{(g)}_s(e^{\lambda_1}+e^{-\lambda_1},\dots, e^{\lambda_s}+e^{-\lambda_s})
={\mathcal F}({\mathcal W}^{(g)}_s)
$$
and ${\mathcal W}^{(g)}_s\in \mathbb Z_{\ge0}[Y^{(g)}_{s} ] $ is given by 
$$
{\mathcal W}^{(g)}_s=
\sum_{(D_0,D_1)\in Y^{(g)}_s}
\widehat b^{(g)}_{D_0,D_1} (D_0,D_1)
$$
where $\widehat b^{(g)}_{D_0,D_1}$ are all non-zero integers.
\end{theorem}

We will provide a proof of this theorem after we have provided recursion relations which completely determine ${\mathcal W}^{(g)}_s\in \mathbb Z_{\ge0}[Y^{(g)}_{s} ] $ from the  initial condition ${\mathcal W}_3^{(0)}=4\,\Cols1\Cols1\Cols1+1\,\Cols1\,\Colsg1\Colsg1$ and
${\mathcal W}^{(1)}_1=1\,\Cols2$.

To this end we consider the ${\mathbb Z}_{\ge 0}$ module ${\mathbb Z}_{\ge0}[Y^{(g)}_{s} ]$.

\begin{definition}
We consider the set $\widetilde Y^{(g)}_{s}$, whose elements are those of $Y^{(g)}_{s}$ together with
the following extra data: The assignment of the label ``1'' to exactly one of the columns of one of the diagrams in the pair. Moreover we require that the label ``1'' is placed at the first column of a given height, if a diagram has more than one columns of the given height.
\end{definition}

We consider the ${\mathbb Z}_{\ge 0}$ module 
${\mathbb Z}_{\ge0}[\widetilde Y^{(g)}_{s} ]$.

\begin{definition}
We define the embedding 
$$S : {\mathbb Z}_{\ge0}[Y^{(g)}_{s} ]\rightarrow {\mathbb Z}_{\ge0}[\widetilde Y^{(g)}_{s} ]$$ 
by letting $S(D_0,D_1)$ be the sum of the elements $\widetilde{(D_0,D_1)}$ (with unit coefficients)
which is the same
pair of Young diagrams $(D_0,D_1)$, but with label ``1'' at all possible places (see examples below). \end{definition}

We observe that  ${\mathcal F}S(D_0,D_1)$ is symmetric in its arguments.

\noindent Let us now consider the operations:

\noindent The \emph{coproduct} operation 
$$\mathrm{CP}: {\mathbb Z}_{\ge0}[Y^{(g)}_{s} ]\to {\mathbb Z}_{\ge0}[\widetilde Y^{(g)}_{s+1} ].$$

\noindent The  \emph{unary operation} $$\mathrm{U}:  {\mathbb Z}_{\ge0}[Y^{(g)}_{s} ]\to {\mathbb Z}_{\ge0}[\widetilde Y^{(g+1)}_{s-1} ].$$

\noindent The \emph{product} operation
$$
 \mathrm{P}:  {\mathbb Z}_{\ge0}[Y^{(g_1)}_{s_1}]\times {\mathbb Z}_{\ge0}[Y^{(g_2)}_{s_2} ]\to 
 {\mathbb Z}_{\ge0}[\widetilde Y^{(g_1+g_2)}_{s_1+s_2-1} ].
 $$

\begin{theorem}
For each $g$ and $s$, one gets that
$$
\mathrm{CP}({\mathcal W}^{(g)}_{s-1})+\mathrm{U}({\mathcal W}^{(g-1)}_{s+1})+\sum_{g_1+g_2=g\atop s_1+s_2=s+1}
\mathrm{P}({\mathcal W}^{(g_1)}_{s_1}\times {\mathcal W}^{(g_2)}_{s_2}) \in S({\mathbb Z}_{\ge0}[Y^{(g)}_{s} ])
$$
and ${\mathcal W}^{(g)}_{s} \in {\mathbb Z}_{\ge0}[Y^{(g)}_{s} ]$ is the unique element such that
\beq
S({\mathcal W}^{(g)}_{s}) = \mathrm{CP}({\mathcal W}^{(g)}_{s-1})+\mathrm{U}({\mathcal W}^{(g-1)}_{s+1})+\sum_{g_1+g_2=g\atop s_1+s_2=s+1}
\mathrm{P}({\mathcal W}^{(g_1)}_{s_1}\times {\mathcal W}^{(g_2)}_{s_2}).
\label{th-7}
\eeq
\end{theorem}

The \emph{coproduct} operation $\mathrm{CP}: {\mathbb Z}_{\ge0}[Y^{(g)}_{s} ]\to {\mathbb Z}_{\ge0}[\widetilde Y^{(g)}_{s+1} ]$ is defined as follows.  

Two columns (labeled ``$1$'' and ``$p$'') are produced out of every column hight (of $D_0$ and $D_1$ separately) in accordance with the following rules:
\begin{itemize}
\item We apply the coproduct operation to exactly one column of every sort (e.g. height and colour). Here and hereafter, we indicate this action by the uparrow symbol standing aside this column: 
\bea
\begin{pspicture}(0.2,1)
\rput(0.1,0.4){$\biggl($}
\end{pspicture}
\Col4
\begin{pspicture}(0.5,1)
\put(0,0){\psline{<->}(0.1,0)(0.1,0.8)}
\rput(0.2,0.4){\makebox(0,0)[lc]{\tiny$k{+}1$}}
\end{pspicture}
\begin{pspicture}(0.5,1)
\rput(0.1,0.4){\makebox(0,0)[lc]{$\biggr)^{\uparrow}$}}
\end{pspicture}
&=&\sum_{m=0}^k2(k-m+1)
\begin{pspicture}(1,1)
\put(0,0){\psline{<->}(0.9,0)(0.9,0.4)}
\rput(0.8,0.2){\makebox(0,0)[rc]{\tiny$m{+}1$}}
\end{pspicture}
\ColOne2
\ColP3
\begin{pspicture}(1,1)
\put(0,0){\psline{<->}(0.1,0)(0.1,0.6)}
\rput(0.2,0.3){\makebox(0,0)[lc]{\tiny$k{-}m{+}1$}}
\end{pspicture}
\nonumber\\
&{}&+
\sum_{m=0}^{k+1}(2k-2m+3)
\biggl[4
\begin{pspicture}(1,1)
\put(0,0){\psline{<->}(0.9,0)(0.9,0.4)}
\rput(0.8,0.2){\makebox(0,0)[rc]{\tiny$m{+}1$}}
\end{pspicture}
\ColOne2
\ColP4
\begin{pspicture}(1.2,1)
\put(0,0){\psline{<->}(0.1,0)(0.1,0.8)}
\rput(0.2,0.4){\makebox(0,0)[lc]{\tiny$k{-}m{+}2$}}
\end{pspicture}
+
\begin{pspicture}(1,1)
\put(0,0){\psline{<->}(0.9,0)(0.9,0.4)}
\rput(0.8,0.2){\makebox(0,0)[rc]{\tiny$m{+}1$}}
\end{pspicture}
\ColgOne2
\ColgP4
\begin{pspicture}(1.2,1)
\put(0,0){\psline{<->}(0.1,0)(0.1,0.8)}
\rput(0.2,0.4){\makebox(0,0)[lc]{\tiny$k{-}m{+}2$}}
\end{pspicture}
\biggr],
\nonumber\\
\begin{pspicture}(0.2,1)
\rput(0.1,0.4){$\biggl($}
\end{pspicture}
\Colg4
\begin{pspicture}(0.5,1)
\put(0,0){\psline{<->}(0.1,0)(0.1,0.8)}
\rput(0.2,0.4){\makebox(0,0)[lc]{\tiny$k{+}1$}}
\end{pspicture}
\begin{pspicture}(0.5,1)
\rput(0.1,0.4){\makebox(0,0)[lc]{$\biggr)^{\uparrow}$}}
\end{pspicture}
&=&\sum_{m=0}^k2(k-m+1)
\begin{pspicture}(1,1)
\put(0,0){\psline{<->}(0.9,0)(0.9,0.4)}
\rput(0.8,0.2){\makebox(0,0)[rc]{\tiny$m{+}1$}}
\end{pspicture}
\ColgOne2
\ColP3
\begin{pspicture}(1,1)
\put(0,0){\psline{<->}(0.1,0)(0.1,0.6)}
\rput(0.2,0.3){\makebox(0,0)[lc]{\tiny$k{-}m{+}1$}}
\end{pspicture}
+
\sum_{m=0}^k (2k-2m+1)
\begin{pspicture}(1,1)
\put(0,0){\psline{<->}(0.9,0)(0.9,0.4)}
\rput(0.8,0.2){\makebox(0,0)[rc]{\tiny$m{+}1$}}
\end{pspicture}
\ColOne2
\ColgP3
\begin{pspicture}(1,1)
\put(0,0){\psline{<->}(0.1,0)(0.1,0.6)}
\rput(0.2,0.3){\makebox(0,0)[lc]{\tiny$k{-}m{+}1$}}
\end{pspicture}
\nonumber\\
&{}&
+
\sum_{m=0}^{k+1} 4(2k-2m+3)
\biggl[
\begin{pspicture}(1,1)
\put(0,0){\psline{<->}(0.9,0)(0.9,0.4)}
\rput(0.8,0.2){\makebox(0,0)[rc]{\tiny$m{+}1$}}
\end{pspicture}
\ColOne2
\ColgP4
\begin{pspicture}(1.2,1)
\put(0,0){\psline{<->}(0.1,0)(0.1,0.8)}
\rput(0.2,0.4){\makebox(0,0)[lc]{\tiny$k{-}m{+}2$}}
\end{pspicture}
+
\begin{pspicture}(1,1)
\put(0,0){\psline{<->}(0.9,0)(0.9,0.4)}
\rput(0.8,0.2){\makebox(0,0)[rc]{\tiny$m{+}1$}}
\end{pspicture}
\ColgOne2
\ColP4
\begin{pspicture}(1.2,1)
\put(0,0){\psline{<->}(0.1,0)(0.1,0.8)}
\rput(0.2,0.4){\makebox(0,0)[lc]{\tiny$k{-}m{+}2$}}
\end{pspicture}
\biggr].
\nonumber
\eea
\item We absorb the thus obtained columns labeled ``$1$'' and ``$p$'' into the pair of Young diagrams (other columns
remain unaltered); among the remaining columns, we have $k\ge0$ columns of the same sort as the column labeled ``$p$''; we then multiply the
resulting diagram by $k+1$ subsequently erasing the label $p$ but retaining the label ``1''.
\end{itemize}

So, applying the coproduct operation we obtain a linear combination of pairs of Young diagrams with positive integer coefficients. 
Exactly one column in each pair is labeled ``1''. 

\begin{example}\label{W04}
We first calculate ${\mathcal W}_4^{(0)}$. Because in this case no product operations are possible, the whole answer is obtained by the
acting of the coproduct operation of ${\mathcal W}_3^{(0)}$. For elements of these Young diagrams we have:
\bea
&{}&(\Cols1)^\uparrow=2\,\ColsOne1\cdot \ColsP1+4\,\ColsOne2\cdot \ColsP1+12\,\ColsOne1\cdot\ColsP2+3\,\ColsgOne1\cdot\ColsgP2
+1\,\ColsgOne2\cdot\ColsgP1 \label{breedOne}\\
&{}&(\Colsg1)^\uparrow=2\,\ColsgOne1\cdot \ColsP1+4\,\ColsgOne2\cdot \ColsP1+1\,\ColsOne1\cdot\ColsgP1+4\,\ColsOne2\cdot\ColsgP1
+12\,\ColsOne1\cdot\ColsgP2+12\,\ColsgOne1\cdot\ColsP2. \label{breedgOne}
\eea
So, for elements of ${\mathcal W}_3^{(0)}$, we obtain (we explicitly segregate the multipliers appearing due to the symmetrisation w.r.t. $p$):
\bea
({\mathcal W}_3^{(0)})^\uparrow &=& 4(\Cols1)^\uparrow\,\Cols1\Cols1+(\Cols1)^\uparrow\,\Colsg1\Colsg1+\Cols1\, (\Colsg1)^\uparrow\,\Colsg1
\nonumber\\
&=&8\cdot 3\ColsOne1\ColsP1\Cols1\Cols1+16\cdot3\,\ColsOne2\ColsP1\Cols1\Cols1+48\, \ColsP2\ColsOne1\Cols1\Cols1+
12\, \Cols1\Cols1\,\ColsgP2\ColsgOne1+4\,\Cols1\Cols1\,\ColsgOne2\ColsgP1\nonumber\\
&{}&\quad+2\,\ColsOne1\ColsP1\,\Colsg1\Colsg1+4\,\ColsOne2\ColsP1\,\Colsg1\Colsg1+12\ColsP2\ColsOne1\,\Colsg1\Colsg1+
3\ColsgP2\ColsgOne1\Colsg1\Colsg1+1\cdot 3\, \ColsgOne2\ColsgP1\Colsg1\Colsg1\nonumber\\
&{}&\quad+2\cdot2\,\ColsP1\Cols1\,\ColsgOne1\Colsg1+4\cdot 2\,\ColsP1\Cols1\,\ColsgOne2\Colsg1
+1\cdot2\,\ColsOne1\Cols1\,\ColsgP1\Colsg1+4\cdot2\, \ColsOne1\Cols2\,\ColsgP1\Colsg1
+12\ColsOne1\Cols1\,\ColsgP2\Colsg1+ 12\ColsP2\Cols1\,\ColsgOne1\Colsg1 \nonumber\\
&=&24\,\ColsOne1\Cols1\Cols1\Cols1 +48\,\Bigl( \ColsOne2\Cols1\Cols1\Cols1+\Cols2\ColsOne1\Cols1\Cols1\Bigr)
+12 \Bigl( \Cols1\Cols1\,\Colsg2\ColsgOne1+\Cols1\Cols1\,\ColsgOne2\Colsg1+\ColsOne1\Cols1\,\Colsg2\Colsg1\Bigr)\nonumber\\
&{}&\quad+12 \Bigl( \ColsOne2\Cols1\,\Colsg1\Colsg1+\Cols2\ColsOne1\,\Colsg1\Colsg1+\Cols2\Cols1\,\ColsgOne1\Colsg1\Bigr)
+3\Bigl( \ColsgOne2\Colsg1\Colsg1\Colsg1+\Colsg2\ColsgOne1\Colsg1\Colsg1\Bigr)
+4\Bigl( \ColsOne1\Cols1\,\Colsg1\Colsg1+\Cols1\Cols1\ColsgOne1\Colsg1\Bigr).\label{W04-ONE}
\eea
We see that we have automatically obtained \emph{symmetrized} expressions w.r.t. $p_1$: every term in brackets contains exactly one
appearance of label ``1'' for every sort of columns thus belonging to 
the image of the mapping $S$. Therefore we get that
\be
{\mathcal W}_4^{(0)}=24\,\Cols1\Cols1\Cols1\Cols1 +48\,\Cols2\Cols1\Cols1\Cols1 +12\, \Cols1\Cols1\,\Colsg2\Colsg1
+12\, \Cols2\Cols1\,\Colsg1\Colsg1 +3\, \Colsg2\Colsg1\Colsg1\Colsg1
+4\,\Cols1\Cols1\,\Colsg1\Colsg1.
\label{W04-TWO}
\ee
We have therefore obtained an expression from ${\mathbb Z}_{\ge0}[Y^{(0)}_{4}]$.
The coefficients in this expression
are exactly $\widehat b_{D_0,D_1}$ for the planar four-backbone case.
\end{example}

The \emph{product} operation produces one column labeled ``1'' (or a linear combination of such columns) out of two columns 
by the following rules:
\be
\begin{pspicture}(0.5,1)
\put(0,0){\psline{<->}(0.4,0)(0.4,0.4)}
\rput(0.3,0.2){\makebox(0,0)[rc]{\tiny$n_1$}}
\end{pspicture}
\underbracket{\Col2\,\,\Col3}
\begin{pspicture}(0.5,1)
\put(0,0){\psline{<->}(0.1,0)(0.1,0.6)}
\rput(0.2,0.3){\makebox(0,0)[lc]{\tiny$n_2$}}
\end{pspicture}
=\ColOne6
\begin{pspicture}(1.2,1)
\put(0,0){\psline{<->}(0.1,0)(0.1,1.2)}
\rput(0.2,0.6){\makebox(0,0)[lc]{\tiny$n_1{+}n_2{+}1$}}
\end{pspicture},
\quad
\begin{pspicture}(0.5,1)
\put(0,0){\psline{<->}(0.4,0)(0.4,0.4)}
\rput(0.3,0.2){\makebox(0,0)[rc]{\tiny$n_1$}}
\end{pspicture}
\underbracket{\Col2\,\,\Colg3}
\begin{pspicture}(0.5,1)
\put(0,0){\psline{<->}(0.1,0)(0.1,0.6)}
\rput(0.2,0.3){\makebox(0,0)[lc]{\tiny$n_2$}}
\end{pspicture}
=\ColgOne6
\begin{pspicture}(1.2,1)
\put(0,0){\psline{<->}(0.1,0)(0.1,1.2)}
\rput(0.2,0.6){\makebox(0,0)[lc]{\tiny$n_1{+}n_2{+}1$}}
\end{pspicture},
\quad
\begin{pspicture}(0.5,1)
\put(0,0){\psline{<->}(0.4,0)(0.4,0.4)}
\rput(0.3,0.2){\makebox(0,0)[rc]{\tiny$n_1$}}
\end{pspicture}
\underbracket{\Colg2\,\,\Colg3}
\begin{pspicture}(0.5,1)
\put(0,0){\psline{<->}(0.1,0)(0.1,0.6)}
\rput(0.2,0.3){\makebox(0,0)[lc]{\tiny$n_2$}}
\end{pspicture}
=4\,\ColOne6
\begin{pspicture}(1.2,1)
\put(0,0){\psline{<->}(0.1,0)(0.1,1.2)}
\rput(0.2,0.6){\makebox(0,0)[lc]{\tiny$n_1{+}n_2{+}1$}}
\end{pspicture}
+\ColOne5
\begin{pspicture}(1.2,1)
\put(0,0){\psline{<->}(0.1,0)(0.1,1)}
\rput(0.2,0.5){\makebox(0,0)[lc]{\tiny$n_1{+}n_2$}}
\end{pspicture}
\label{1}
\ee

We have two cases.

\noindent{\bf 1.}\  The first case is where we do the product 
\emph{inside} the same pair of diagrams $(D_0,D_1)$, 
 $$\mathrm{U}:  {\mathbb Z}_{\ge0}[Y^{(g)}_{s} ]\to {\mathbb Z}_{\ge0}[\widetilde Y^{(g+1)}_{s-1} ].$$
 In this case, we must take 
all possible (pairwise) products between different types of columns (one product operation per every pair of types) as well as
products inside the same type (if we have more than one column of this type in $D_0$ or in $D_1$).
The additional factors are:
\begin{itemize}
\item we have a factor of two if we make the product between different types of columns;
\item we have an additional factor of two if we make a product in a term of ${\mathcal W}^{(g)}_s$ with $s>2$,
i.e., if the result of U when acting on $(D_0,D_1)$ contains more than one column.
\end{itemize}

\noindent{\bf 2.}\ The second case is where we do the product of \emph{two different} pairs of Young
diagrams  $(D_0,D_1)\in Y^{(g_1)}_{s_1}$ and
$(D'_0,D'_1)\in Y^{(g_2)}_{s_2}$, 
 $$
 \mathrm{P}:  {\mathbb Z}_{\ge0}[Y^{(g_1)}_{s_1}]\times {\mathbb Z}_{\ge0}[Y^{(g_2)}_{s_2} ]\to 
 {\mathbb Z}_{\ge0}[\widetilde Y^{(g_1+g_2)}_{s_1+s_2-1} ].
 $$
In this case, we must make 
all possible products between all column types in the first pair and in the second pair
(one product per every pair of types from different pairs of diagrams) unless $(D_0,D_1)=(D_0',D_1')$; in the latter
case we take into account every type of pairings between entries of the diagrams $(D_0\,D_1)$ only once. 
We then take the union of two above pairs of Young diagrams; as the result, we
obtain a linear combination of Young diagrams of the form
\beq
\underbrace{\Col5\Col5\Col5\Col5\Col5}_{\hbox{\tiny $d_k$}}\!\underbrace{\Col5\Col5\Col5\Col5}_{\hbox{\tiny $d'_k$}}\cdots
\!\ColOne4\!
\underbrace{\Col4\Col4\Col4\Col4\Col4}_{\hbox{\tiny $d_l$}}\!\underbrace{\Col4\Col4\Col4\Col4}_{\hbox{\tiny $d'_l$}}\cdots
\underbrace{\Col2\Col2\Col2\Col2}_{\hbox{\tiny $d_2$}}\!\underbrace{\Col2\Col2\Col2}_{\hbox{\tiny $d'_2$}}
\underbrace{\Col1\Col1\Col1\Col1\Col1}_{\hbox{\tiny $d_1$}}\!\underbrace{\Col1\Col1\Col1}_{\hbox{\tiny $d'_1$}}\,
\underbrace{\Colg4\Colg4\Colg4\Colg4\Colg4}_{\hbox{\tiny $t_r$}}\!\underbrace{\Colg4\Colg4\Colg4\Colg4}_{\hbox{\tiny $t'_r$}}\cdots
\underbrace{\Colg1\Colg1\Colg1\Colg1\Colg1}_{\hbox{\tiny $t_1$}}\!\underbrace{\Colg1\Colg1\Colg1}_{\hbox{\tiny $t'_1$}},
\eeq
where we have exactly one column labeled ``1'' and in every term $d_j$ columns come from $D_0$, $d'_j$ columns come 
from $D'_0$ and, correspondingly, $t_i$ columns come from $D_1$ and $t'_i$ columns come from $D'_1$. 

The combinatorial factors are:
\begin{itemize}
\item we multiply the obtained Young diargam by the product of binomial factors:
$$
\prod_{j=1}^k \binom{d_j+d'_j}{d_j}\prod_{i=1}^r \binom{t_i+t'_i}{t_i};
$$
\item we multiply by a factor of two if we make the product between different types of columns and/or if we
make a product between two different diagrams (i.e., if $D_0\ne D'_0$ and/or $D_1\ne D'_1$)
(in other words, the only situation when we do not have this factor is when we evaluate the product between two equal
Young diagrams, $D_0=D'_0$ and $D_1=D'_1$ and we make a product of terms of the same sort in these two diagrams);
\item we multiply by an additional factor of two 
if the result of product of Young diagrams $(D_0,D_1)$ and $(D'_0,D'_1)$ contains more than one column,
that is, if $s_1=l(D_0)+l(D_1)>1$ and/or $s_2=l(D'_0)+l(D'_1)>1$ (in other words, the only situation when we do not have this factor is when
$s_1=s_2=1$).
\end{itemize}

\begin{example}
We next calculate ${\mathcal W}_2^{(1)}$ (in the third line, we explicitly indicate the combinatorial factors due to the product process):
\bea
{\mathcal W}_2^{(1)}&=&({\mathcal W}_1^{(1)})^\uparrow+\underbracket{{\mathcal W}_3^{(0)}}
=\bigl(\Cols2\bigr)^\uparrow+4\underbracket{\Cols1\Cols1}\Cols1+
\underbracket{\Cols1\Colsg1}\Colsg1+\Cols1\,\underbracket{\Colsg1\Colsg1}\nonumber\\
&=&4\,\Cols2\ColsOne1+12\,\ColsOne2\Cols2+2\,\ColsOne2\Cols1+4\,\ColsOne3\Cols1+20\,\Cols3\ColsOne1+5\,\Colsg3\ColsgOne1
+3\,\ColsgOne2\Colsg2+\ColsgOne3\Colsg1 \nonumber\\
&{}&\quad+4\cdot 2\, \ColsOne3\Cols1+1\cdot 4\,\ColsgOne3\Colsg1+4\cdot 2\, \ColsOne3\Cols1
+1\cdot 2\, \ColsOne2\Cols1 \nonumber\\
&=&4\Bigl(\Cols2\ColsOne1+\ColsOne2\Cols1 \Bigr)+12\,\ColsOne2\Cols2 +20\Bigl(\Cols3\ColsOne1+\ColsOne3\Cols1 \Bigr)
+5\Bigl(\Colsg3\ColsgOne1+\ColsgOne3\Colsg1 \Bigr)+3\,\ColsgOne2\Colsg2\nonumber\\
&=&4\,\Cols2\Cols1+12\,\Cols2\Cols2+20\,\Cols3\Cols1+5\,\Colsg3\Colsg1+3\,\Colsg2\Colsg2.\label{W21-ONE}
\eea
Having this expression and ${\mathcal W}_1^{(1)}$, we can now calculate ${\mathcal W}_1^{(2)}$:
\bea
{\mathcal W}_1^{(2)}&=&\underbracket{{\mathcal W}_2^{(1)}}+\underbracket{{\mathcal W}_1^{(1)}\times {\mathcal W}_1^{(1)}}=
4\,\underbracket{\Cols2\Cols1}+12\,\underbracket{\Cols2\Cols2}+20\,\underbracket{\Cols3\Cols1}+5\,
\underbracket{\Colsg3\Colsg1}+3\,\underbracket{\Colsg2\Colsg2}+\underbracket{\Cols2\times \Cols2}\nonumber\\
&=&4\cdot2\,\ColOne4+12\,\ColOne5+20\cdot 2\,\ColOne5+5\cdot 2\Bigl(4\,\ColOne5+\ColOne4\Bigr)
+3\Bigl(4\,\ColOne5+\ColOne4\Bigr)+1\cdot 1\,\ColOne5\nonumber\\
&=& 3\cdot 5\cdot 7\,\Col5+3\cdot 7\,\Col4.\label{W12-ONE}
\eea
The same answer follows from the Harer--Zagier recursion relation: $b_1^{(2)}=3\cdot5\cdot 7$, 
$b_0^{(2)}=3\cdot 7$.
\end{example}

\begin{example}
The first example in which we have all three above operations is calculating ${\mathcal W}_3^{(1)}$:
\be
{\mathcal W}_3^{(1)}=\bigl({\mathcal W}_2^{(1)}\bigr)^\uparrow + \underbracket{{\mathcal W}_4^{(0)}}
+\underbracket{{\mathcal W}_3^{(0)}\times {\mathcal W}_1^{(1)}}.
\label{W31-ONE}
\ee
Here, the first term (with coproduct) contains 67 summands, the second contains 21 summands, and the third contains only three
summands, and performing the summation we again obtain the result that is totally 
symmetric in all $p_i$ including $p_1$ and reads
\bea
{\mathcal W}_3^{(1)}&=&24\,\Col2\Col1\Col1+192\,\Col2\Col2\Col1+240\,\Col3\Col1\Col1+288\,\Col2\Col2\Col2+480\,\Col3\Col2\Col1
+560\,\Col4\Col1\Col1+30\,\Col1\,\Colg3\Colg1+18\,\Col1\Colg2\Colg2+24\,\Col2\,\Colg2\Colg1\nonumber\\
&{}&+120\,\Col2\,\Colg3\Colg1+72\,\Col2\,\Colg2\Colg2+140\,\Col1\,\Colg4\Colg1+120\,\Col1\,\Colg3\Colg2
+120\,\Col3\,\Colg2\Colg1+30\,\Col3\,\Colg1\Colg1+140\,\Col4\,\Colg1\Colg1.\label{W31-TWO}
\eea
One more example is ${\mathcal W}_5^{(0)}$:
\be
{\mathcal W}_5^{(0)}=\bigl({\mathcal W}_4^{(0)}\bigr)^\uparrow +\underbracket{{\mathcal W}_3^{(0)}\times {\mathcal W}_3^{(0)}}.
\label{W50-ONE}
\ee
The first term comprises 85 summands and the second term comprises seven summands presenting below for
clarifying the symmetry coefficients count (we omit unit binomial coefficients)
\bea
\underbracket{{\mathcal W}_3^{(0)}\times {\mathcal W}_3^{(0)}}&=&16\underbracket{\Cols1\Cols1\Cols1\times\Cols1}\!\Cols1\Cols1
+4\underbracket{\Cols1\Cols1\Cols1\times\Cols1}\,\Colsg1\Colsg1+4\underbracket{\Cols1\Cols1\Cols1\times\Cols1\,\Colsg1}\!\Colsg1
+\underbracket{\Cols1\,\Colsg1\Colsg1\times\Cols1}\,\Colsg1\Colsg1+
\underbracket{\Cols1\,\Colsg1\Colsg1\times\Cols1\,\Colsg1}\!\Colsg1+\Cols1\,\underbracket{\Colsg1\Colsg1\times\Cols1\,\Colsg1}\!\Colsg1
\nonumber\\
&=&16\cdot2\cdot{\hbox{\small$\binom{4}{2}$}}\,\ColsOne3\Cols1\Cols1\Cols1\Cols1+4\cdot2\cdot2\,\ColsOne3\Cols1\Cols1\,\Colsg1\Colsg1
+4\cdot 2\cdot 2\cdot {\hbox{\small$\binom{3}{1}$}}\,\Cols1\Cols1\Cols1\,\ColsgOne3\Colsg1\nonumber\\
&{}&\quad
+2\cdot{\hbox{\small$\binom{4}{2}$}}\,\ColsOne3\,\Colsg1\Colsg1\Colsg1\Colsg1
+2\cdot 2\cdot{\hbox{\small$\binom{3}{1}$}}\,\Cols1\,\ColsgOne3\Colsg1\Colsg1\Colsg1
+2\cdot{\hbox{\small$\binom{2}{1}\cdot\binom{2}{1}$}}\,\Cols1\Cols1\Bigl(4\ColsgOne3+\ColsgOne2\Bigr)\Colsg1\Colsg1.
\nonumber
\eea
Here, because we apply the product operation to identical objects, we take into account every type of product only once, but if we make a product of
different entries or different types of columns inside the same entry, we have to multiply by two. One factor of two is always present because the result contains more than one column.

The sum in (\ref{W50-ONE}) is totally
symmetric in all $p_i$ including $p_1$ and reads
\bea
{\mathcal W}_5^{(0)}&=&192\,\Cols1\Cols1\Cols1\Cols1\Cols1+768\,\Cols2\Cols1\Cols1\Cols1\Cols1+1152\,\Cols2\Cols2\Cols1\Cols1\Cols1
+960\,\Cols3\Cols1\Cols1\Cols1\Cols1+144\,\Cols1\Cols1\Cols1\,\Colsg2\Colsg1+240\,\Cols1\Cols1\Cols1\,\Colsg3\Colsg1\nonumber\\
&{}&+288\,\Cols2\Cols1\Cols1\,\Colsg2\Colsg1+288\,\Cols1\Cols1\Cols1\,\Colsg2\Colsg2+144\,\Cols2\Cols1\Cols1\,\Colsg1\Colsg1
+288\,\Cols2\Cols1\Cols1\,\Colsg1\Colsg1+240\,\Cols3\Cols1\Cols1\,\Colsg1\Colsg1+24\,\Cols1\Cols1\Cols1\,\Colsg1\Colsg1\nonumber\\
&{}&+72\,\Cols1\,\Colsg2\Colsg2\Colsg1\Colsg1 +60\,\Cols1\,\Colsg3\Colsg1\Colsg1\Colsg1+18\,\Cols1\,\Colsg2\Colsg1\Colsg1\Colsg1
+72\,\Cols2\,\Colsg2\Colsg1\Colsg1\Colsg1+12\,\Cols2\,\Colsg1\Colsg1\Colsg1\Colsg1+60\,\Cols3\,\Colsg1\Colsg1\Colsg1\Colsg1
\label{W50-TWO}
\eea
\end{example}

We see that all the coefficients in the coproduct and product relations are positive integers, so the result is always integral and positive.

\noindent
{\bf Proof of Theorem 6 and 7.\ }
Recurrent relations  (\ref{recursion}) are fundamental relations of the topological recursion: it follows from the results of \cite{ChEy} and \cite{CEO} that, for any spectral curve, the $s$-differentials $W^{(g)}_s(p_1,\dots, p_s)$ obtained as a result of successive application of this relation have poles only at zeros of $dx$ and are totally symmetric w.r.t. permutations of all their arguments $p_1,\dots, p_s$. 

We demonstrate now that the operations P, U, and CP on the set of Young diagrams represent recurrent relations (\ref{recursion}).
The mapping (\ref{hom-F}) is obviously invertible on the set of $W^{(g)}_s(p_1,\dots, p_s)$ that have poles of finite order only at $\lambda=0,i\pi$ and are skew-symmetric w.r.t. $\lambda_i\to -\lambda_i$. Note that this inverse mapping exists and is uniquely defined also for non-symmetric products of basis functions. This means that provided we have represented the terms $\mathcal W_s^{(g)}$ in the both sides of relation (\ref{th-7}) using the mapping (\ref{hom-F}) and provided we have demonstrated that, upon this mapping, the relation (\ref{th-7}) follows from (\ref{recursion}), we obtain that (i) the left-hand side of (\ref{recursion}) upon the inverse mapping ${\mathcal F}^{-1}$ is a finite linear combination of pairs of Young diagrams and (ii) this expression is in the image of $S$ being totally symmetric w.r.t. all its arguments.

Recurrent relations described by (\ref{recursion}) (or graphically in Fig.~\ref{fi:recursion}) can be rewritten as two operations on the
basic one-differentials $s_{k,0}(\lambda)dx$ and $s_{k,1}(\lambda)dx$ (where $x=e^\lambda+e^{-\lambda}$): we are going to demonstrate that
those are the ``product'' and ``coproduct'' operations presented above on the set of Young diagrams.

The product operation occurs in the second and
third terms in  (\ref{recursion}). It suffices to define it on the set of basis one-differentials and continue by bi-linearity to products of these one-differentials constituting
$W^{(g)}_s(e^{\lambda_1}+e^{-\lambda_1},\dots, e^{\lambda_s}+e^{-\lambda_s})$. On the level of functions,
this operation produces a linear combination of basis functions $s_{k,\beta}(\lambda_1)$ out of two basis functions,
$s_{k_1,\beta_1}(\lambda_{r_1})$ and $s_{k_2,\beta_2}(\lambda_{r_2})$ ``forgetting'' the initial arguments $\lambda_{r_i}$. We denote
this operations by a standard ``pairing'' symbol and define it to be the following integral (in which $x=e^\lambda+e^{-\lambda}$ and $q=e^\lambda$):
\be
\hbox{``product'':}\quad \quad \underbracket {s_{k_1,\beta_1}(\lambda_{r_1})dx_{r_1}\ \  s_{k_2,\beta_2}(\lambda_{r_2})dx_{r_2}}:=
\sum_{\mathrm{res\,}dx=0} K(p_1,q)s_{k_1,\beta_1}(\lambda)s_{k_2,\beta_2}(\lambda)dq.
\label{contr1}
\ee
Recalling that $dx=(e^{\lambda}-e^{-\lambda})d\lambda$ and that, instead of evaluating this integral by residues at the branch points we
can evaluate it at its only simple pole $p_1=q$ outside the branch points, we obtain with accounting for explicit form (\ref{sj}) of the basic vectors
\bea
&{}&
\sum_{\mathrm{res\,}dx=0} K(p_1,q)s_{k_1,\beta_1}(\lambda)s_{k_2,\beta_2}(\lambda)dq=-\res_{p_1=q}K(p_1,q)s_{k_1,\beta_1}(\lambda)s_{k_2,\beta_2}(\lambda)dq\nonumber\\
&{}&\quad=\frac{(e^{\lambda_1}+e^{-\lambda_1})^{\beta_1+\beta_2}}{(e^{\lambda_1}-e^{-\lambda_1})^{6+2k_1+2k_2}}d\lambda_1
=\frac{(e^{\lambda_1}+e^{-\lambda_1})^{\beta_1+\beta_2}}{(e^{\lambda_1}-e^{-\lambda_1})^{7+2k_1+2k_2}}dx_1,\nonumber
\eea
so, recalling that $(e^{\lambda_1}+e^{-\lambda_1})^2=(e^{\lambda_1}-e^{-\lambda_1})^2+4$, we obtain the following rule for the product operation:
\be
\underbracket {s_{k_1,\beta_1}(\lambda_{r_1})dx_{r_1}\ \  s_{k_2,\beta_2}(\lambda_{r_2})dx_{r_2}}=
\left\{\begin{array}{ll} s_{k_1+k_2+2,\beta_1+\beta_2}(\lambda_1)dx_1& \beta_1+\beta_2< 2,\\
s_{k_1+k_2+1,0}(\lambda_1)dx_1+4s_{k_1+k_2+2,0}(\lambda_1)dx_1& \beta_1=\beta_2=1.
\end{array}
\right.
\label{contr2}
\ee
On the level of Young diagrams, this operation generates the product operation P when the above two basis vectors belong to different pairs of Young diagrams and it generates the unary operation U when these vectors belong to the same pair of Young diagrams. We therefore use the same symbols to denote these operations either on Young diagrams or on basis vectors.

The second operation we need is the ``coproduct'' operation, which we encounter in the first term on the right-hand side of (\ref{recursion}). This operation
produces a term bilinear in $s_{k_1,\beta_1}(\lambda_1)dx_1$ and $s_{k_2,\beta_2}(\lambda_p)dx_p$ out of a basis one-differential
 $s_{k,\beta}(\lambda_r)dx_r$ forgetting the argument $\lambda_r$, is denoted by uparrow aside the symbol of this differential, and
 is to be continued by linearity to products of these basic differentials. It is given by the following integral (where $x=e^\lambda+e^{-\lambda}$ and $q=e^\lambda$):
\be
\hbox{``coproduct'':}\quad  \Bigl(s_{k,\beta}(\lambda_r)dx_r\Bigr)^{\uparrow}:=\sum_{\mathrm{res\,}dx=0} K(p_1,q)\bigl[ B(p,q)+B(\overline p,q)\bigr]
s_{k,\beta}(\lambda),
\label{breed1}
\ee
where we can again do the integration by residues at $q=p_1$ and $q=p$ (for the term with $B(p,q)$) 
and at $q=p_1$ and $q=\overline p$ (for the term with $B(\overline p,q)$). The calculations involve combinatorics of geometric progression type but 
are otherwise straightforward. Two cases, $\beta=0$ and $\beta=1$, are rather different, so two integrations yield two
cases of the ``coproduct'' operation (here $q=e^\lambda$, $x=e^{\lambda}+e^{-\lambda}$, 
$p=e^{\lambda_p}$, $\overline p=e^{-\lambda_p}$, and $x_p=e^{\lambda_p}+e^{-\lambda_p}$):
\bea
 \Bigl(s_{k,0}(\lambda_r)dx_r\Bigr)^{\uparrow}&=&\sum_{\mathrm{res\,}dx=0} K(p_1,q)\bigl[ B(p,q)+B(\overline p,q)\bigr]
s_{k,1}(\lambda)\nonumber\\
&=&\sum_{m=0}^k (2+2k-2m)s_{m,0}(\lambda_1)s_{k-m,0}(\lambda_p)dx_1 dx_p\nonumber\\
&{}&\hskip-1.2cm+\sum_{m=0}^{k+1} (3+2k-2m)\Bigl[4s_{m,0}(\lambda_1)s_{k+1-m,0}(\lambda_p)
+s_{m,1}(\lambda_1)s_{k+1-m,1}(\lambda_p)\Bigr]dx_1 dx_p
\label{breed2}
\eea
and
\bea
&{}& \Bigl(s_{k,1}(\lambda_r)dx_r\Bigr)^{\uparrow}=\sum_{\mathrm{res\,}dx=0} K(p_1,q)\bigl[ B(p,q)+B(\overline p,q)\bigr]
s_{k,1}(\lambda)\nonumber\\
&{}&=\sum_{m=0}^k (2+2k-2m)s_{m,1}(\lambda_1)s_{k-m,0}(\lambda_p)dx_1 dx_p
+\sum_{m=0}^k (1+2k-2m)s_{m,0}(\lambda_1)s_{k-m,1}(\lambda_p)dx_1 dx_p
\nonumber\\
&{}&\qquad +\sum_{m=0}^{k+1} 4(3+2k-2m) \Bigl[s_{m,0}(\lambda_1)s_{k+1-m,1}(\lambda_p)
+s_{m,1}(\lambda_1)s_{k+1-m,0}(\lambda_p)\Bigr]dx_1 dx_p
\label{breed3}
\eea
Upon the inverse mapping ${\mathcal F}^{-1}$, this coproduct operation generates the CP operation on the level of Young diagrams.

The two correlation functions we need to commence the recursion procedure are
\bea
&{}&W_3^{(0)}(x_1,x_2,x_3)=4s_{0,0}(\lambda_1)s_{0,0}(\lambda_2)s_{0,0}(\lambda_3)dx_1dx_2dx_3\nonumber\\
&{}&\!\!\!\!\!\!\!\!+\bigl[s_{0,1}(\lambda_1)s_{0,1}(\lambda_2)s_{0,0}(\lambda_3)+
s_{0,1}(\lambda_1)s_{0,0}(\lambda_2)s_{0,1}(\lambda_3)+
s_{0,0}(\lambda_1)s_{0,1}(\lambda_2)s_{0,1}(\lambda_3)\bigr]dx_1dx_2dx_3
\label{W30-GUE}
\eea
and
\bea
W^{(1)}_1(x)=s_{1,0}(\lambda)dx.
\label{W11-GUE}
\eea

Both the product and coproduct operations are closed on the linear space of ${\mathcal F}^{-1}(\widetilde{D_0,D_1})$; besides that, for a modules
${\mathbb Z}_{\ge0}(\mathop{\otimes}_{g,s}Y_{s}^{(g)})$, the result of application of every such operation lies in 
${\mathbb Z}_{\ge0}(\mathop{\otimes}_{g,s}{\mathcal F}(\widetilde Y_{s}^{(g)}))$.

Moreover, due to the lemma in \cite{ChEy}, the result of joint application of coproduct and product operations in  (\ref{recursion})
is automatically symmetric w.r.t. permutations of all arguments including $p_1$, so the result of joint application
of the operations CP, U, and P  in formula (\ref{th-7})
lies in the image $S(Y^{(g)}_s)$ in $\widetilde Y^{(g)}_s$, which we can naturally identify with
$Y^{(g)}_s$ itself. This completes the proof of theorems.$.\qquad\square$

\section{The one-backbone case}
\label{s:HZ}
\setcounter{equation}{0}

\subsection{The Harer--Zagier recursion and the graph decomposition from Sec.~\ref{s:KPMM}}

In the one-backbone case, we have the representation (\ref{W1g-B}) and the alternative representation
\beq
W^{(g)}_1(e^{\lambda}+e^{-\lambda})=\sum_{r=0}^{3g-2}
(-1)^r \frac{\varkappa_{g,1,r}}{2^{d-r}(d-r)!}\frac{1}{e^{\lambda}-e^{-\lambda}}
\left(\frac{\partial}{\partial \lambda}\right)^{2d-2r+1}\frac{2}{e^{2\lambda}-1},\quad d=3g-2,
\label{one-bone}
\eeq
where on the base of reasonings related to stratification of closed moduli spaces,
$\varkappa_{g,1,r}$ are (conjecturally positive) rational numbers, $\varkappa_{g,1,0}=\<\tau_{3g-2}\>_g$.

For the coefficients $b^{(g)}_i$ of (\ref{W1g-B}) based on the Harer and Zagier recurrent formula \cite{HZ},
 we have obtained the recurrence relation (also found in  \cite{HT})

\begin{proposition}
\label{th5} {\rm\cite{ACNP}}
The coefficients $b^{(g)}_k$ from~(\ref{W1g-B}) satisfy the three-term recurrence relation:
\beq
\label{recurrence}
(4g+2k+6)b^{(g+1)}_k=(4g+2k+1)(4g+2k+3)\Bigl[(4g+2k+2)b^{(g)}_k+4(4g+2k-1)b^{(g)}_{k-1}\Bigr].
\eeq
All these coefficients are positive integers.
\end{proposition}

(Of course, the positive integrality of $b^{(g)}_k$ is a particular case of the general Theorem~\ref{thm-integral}.)

In \cite{ACNP}, we used recursion (\ref{recurrence}) to develop a method allowing determining $b^{(g)}_{g-1-k}$ for any fixed
$k\ge 0$ and for all $g$. For example, we have just two-term relations for the boundary coefficients 
\bea
(4g+6)b_0^{(g+1)}&=&(4g-1)(4g+3)(4g+2)b_0^{(g)},\nonumber\\
(6g+6)b_g^{(g+1)}&=&4(6g+1)(6g+3)(6g-1)b_{g-1}^{(g)},\nonumber
\eea
which immediately give
\beq
\label{b0-bg}
b_{g-1}^{(g)}=\frac{2^{g-1}\,(6g-3)!!}{3^g\, g!},\qquad b_{0}^{(g)}=\frac{(4g)!}{8^g\,g!\,(2g+1)!!}.
\eeq

Substituting
$b_{g-1}^{(g)}$ into (\ref{one-bone}) and evaluating the leading term ($r=0$) we obtain the
highest Kontsevich coefficient $\varkappa_{g,1,0}=\left\langle \tau_{3g-2}\right\rangle_g=\frac{1}{2^{3g}\,3^g\,g!}$. 

Solving recursion (\ref{recurrence}) for the first subleading term, we have obtained
\beq
b^{(g)}_{g-2}=\frac 15 \frac{2^{g-2}\,(6g-5)!!}{3^{g-2}\,(g-2)!}, \quad\hbox{or}\quad
\varkappa_{g,1,1}=\frac 15[12g^2-7g+5]\varkappa_{g,1,0},\quad g\ge 2.
\label{bgg-2}
\eeq
For the next term, we have
\beq
\label{b-g-3}
b^{(g)}_{g-3}=\frac{(2g-1)\,2^{g-3}\,(6g-7)!!}{5^2\,3^{g-3}\,(g-3)!}-\frac{7\,2^{g-3}\,(6g-7)!!}{10\,(3g-2)!!!},
\ \hbox{where}\ (3g-2)!!!\equiv \prod_{k=3}^g(3k-2),
\eeq
etc. The complete multi-step procedure was described in \cite{ACNP}.

We can alternatively derive $b^{(g)}_{g-2}$ from the graph representation of Lemma \ref{lm:graph}.
For this, it suffices to take only the part with the times $T^+_{2k}$. The highest term for genus $g$ is $\<\tau_{3g-2}\>_gT^+_{6g-4}$

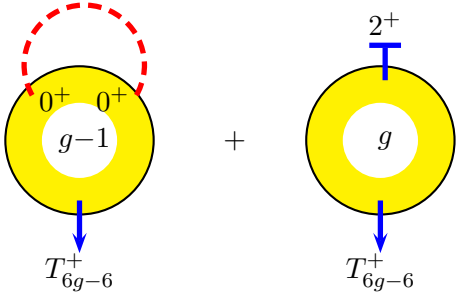
\begin{figure}[h]
{\psset{unit=1}
\begin{pspicture}(-5,-2)(3,2)
\newcommand{\PATTERN}{%
{\psset{unit=1}
\pscircle[fillstyle=solid,fillcolor=yellow](0,0){1}
\pscircle[linecolor=white,fillstyle=solid,fillcolor=white](0,0){0.5}
\psline[linewidth=2pt,linecolor=blue]{->}(0,-0.8)(0,-1.5)
\rput(0,-1.5){\makebox(0,0)[ct]{$T^+_{6g-6}$}}
}
}
\rput(-2,0){\PATTERN}
\rput(-2,0){\makebox(0,0)[cc]{$g{-}1$}}
\psarc[linecolor=red,linestyle=dashed,linewidth=2pt](-2,1){0.8}{-30}{210}
\rput(-2.6,0.55){\makebox(0,0)[lc]{$0^+$}}
\rput(-1.4,0.55){\makebox(0,0)[rc]{$0^+$}}
\rput(0,0){\makebox(0,0)[cc]{$+$}}
\rput(2,0){\PATTERN}
\rput(2,0){\makebox(0,0)[cc]{$g$}}
\psline[linewidth=2pt,linecolor=blue]{-|}(2,0.8)(2,1.3)
\rput(2,1.4){\makebox(0,0)[cb]{$2^+$}}
\end{pspicture} }
\caption{\small The two diagrams contributing to $b^{(g)}_{g-2}$.}
\label{fi:firstcorrection}
\end{figure}

Following Lemma~\ref{lm:graph}, the first-order
correction, or the coefficient of $T^+_{6g-6}$, comes only from two terms: from the graph with
one vertex and one internal edge with endpoint markings $(0,0)$ and from the graph with one vertex and one half-edge
with marking 2 (see Fig.~\ref{fi:firstcorrection}): the corresponding coefficient is then
\beq
\frac{B_2}{4}\<\tau_{3g-3}\tau_0\tau_0\>_{g-1}+\frac{2^3}{5!}\<\tau_{3g-3}\tau_2\>_g,
\label{bg-g-2-rec}
\eeq
and we need only to know the corresponding intersection indices. Whereas
$\<\tau_{3g-3}\tau_0\tau_0\>_{g-1}=\<\tau_{3g-5}\>_{g-1}$, in \cite{ACNP} we have calculated
the intersection index
$\<\tau_{3g-3}\tau_2\>_g$ using the Virasoro conditions for the Kontsevich matrix model; the result is
\beq
\<\tau_2\tau_{3g-3}\>_{g}=\frac15 [12g(g-1)+5]\<\tau_{3g-2}\>_g,\quad g\ge2.
\label{index}
\eeq
Using formula (\ref{index}) and that $B_2=1/24$, we obtain that the
coefficient of $T^{+}_{6g-6}$ is
\beq
\frac15 [12g^2-7g+5]
\label{coefficient}
\eeq
in full agreement with (\ref{bgg-2}).

Below we present the third calculation of the same quantity using the explicit diagram counting.

\subsection{$2$-cycles and the recursion for $b^{(g)}_{g-1}$ and  $b^{(g)}_{g-2}$ terms}\label{ss:blowing}

\subsubsection{Contracting edges in genus-$g$ graphs}

We now find $b^{(g)}_{g-2}$ using the explicit fat graph counting. For this, we consider the set of shapes
with one boundary component and one marked edge. We let $\Gamma^{(g)}$ denote the sets of combinatorial types of
the corresponding shapes of genus $g$ and let $V^{(g)}$ denote cardinalities of these sets.

We first consider the procedure of edge contraction in the genus $g$
graphs. We let $\Gamma^{(g)}_{q;3{-}3}$ denote the set of genus-$g$
shapes with the marked edge with all vertices having valence three {\em and with $q$ 2-cycles}
(all these 2-cycles are of the form as in the rightmost diagram in
Fig.~\ref{fi:handle}). We let $V^{(g)}_{q;3{-}3}$ denote the number
of such diagrams. We let $\Gamma^{(g)}_{4,3{-}3}$,
$\Gamma^{(g)}_{4,4,3{-}3}$, and $\Gamma^{(g)}_{5,3{-}3}$ denote the
respective sets of of genus-$g$ shapes with the marked edge and with one four-valent
vertex, two four-valent vertices, and one five-valent vertex and
with all other vertices having valence three. The numbers of the
corresponding shapes are $V^{(g)}_{4,3{-}3}$,
$V^{(g)}_{4,4,3{-}3}$, and $V^{(g)}_{5,3{-}3}$.

We now consider the contraction process. We never contract the
marked edge corresponding to the ends of the backbone and can contract any other edge in any graph from
$\Gamma^{(g)}_{q;3{-}3}$ (there are $6g-4$ contractible edges in
total) every time obtaining a graph from $\Gamma^{(g)}_{4,3{-}3}$.
Vice versa, every graph from $\Gamma^{(g)}_{4,3{-}3}$ can be
obtained from two graphs in $\Gamma^{(g)}_{q;3{-}3}$; we therefore
have the equality
$$
(6g-4)\sum_{q=0}^{\mathrm{max}}V^{(g)}_{q;3{-}3}=2V^{(g)}_{4,3{-}3}=(6g-4)b^{(g)}_{g-1}.
$$

A more interesting situation occurs when we want to contract two edges. We have three possible outcomes:
\begin{itemize}
\item[1] when we contract two disjoint edges we obtain a graph from $V^{(g)}_{4,4,3{-}3}$;
\item[2] when we contract two edges with incidence one we obtain a graph from $V^{(g)}_{5,3{-}3}$;
\item[3] we do not allow contracting two edges with incidence two (which therefore constitute a 2-loop).
\end{itemize}

We consider the first case first. The total number of disjoint pair of edges is
\beq
\frac12(6g-4)(6g-5)-\hbox{\# of incident pairs of edges}.
\eeq
The number of edges of incidence one and two can be easily counted: this is three times 
the number of vertices minus 4
because of the marked edge minus twice the number of 2-loops in a graph from $V^{(g)}_{q,3{-}3}$, i.e.,
$$
3(4g-2)-4-2q;
$$
the number of pairs of incidence two is obviously $q$. Then the total number of nonincident pairs can be easily counted to be
$$
(3g-4)(6g-5)+q.
$$
Note that from each such pair we produce a graph in
$V^{(g)}_{4,4,3{-}3}$, and each graph from $V^{(g)}_{4,4,3{-}3}$ can
be produced exactly in four ways from the graphs from
$V^{(g)}_{q;3{-}3}$ with some $q$ (it might be the same graph from
$V^{(g)}_{q;3{-}3}$ that produces a graph from
$V^{(g)}_{4,4,3{-}3}$, we then count this case with the
corresponding multiplicity. The resulting relation reads
\beq
\label{3-3->44}
\sum_{q=0}^{\mathrm{max}}\bigl[(3g-4)(6g-5)+q\bigr]V^{(g)}_{q;3{-}3}=4 V^{(g)}_{4,4,3{-}3}.
\eeq
Analogously, each graph from $V^{(g)}_{5,3{-}3}$ can be obtained by
contracting two edges with incidence one by exactly five ways from
graphs in $V^{(g)}_{q;3{-}3}$, that is, we obtain that
\beq
\label{3-3->5}
\sum_{q=0}^{\mathrm{max}}\bigl[12g-10-2q\bigr]V^{(g)}_{q;3{-}3}=5 V^{(g)}_{5,3{-}3}.
\eeq
The total number of diagrams with $6g-6$ nonmarked edges is
precisely the sum of $V^{(g)}_{4,4,3{-}3}$ and $V^{(g)}_{5,3{-}3}$,
and it is given by a combination of $b$ factors, so we obtain
\beq
\label{6g-6}
V^{(g)}_{4,4,3{-}3}+V^{(g)}_{5,3{-}3}=b^{(g)}_{g-2}+\frac{(3g-2)(3g-3)}{2}b^{(g)}_{g-1},
\eeq
and we have three above equations on three unknowns $V^{(g)}_{4,4,3{-}3}$, $V^{(g)}_{5,3{-}3}$, and
$\sum_{q=1}^{\mathrm{max}}qV^{(g)}_{q;3{-}3}$. The solution reads
\bea
V^{(g)}_{4,4,3{-}3}&=&\frac14\Bigl[(3g-4)(6g-5)b^{(g)}_{g-1}+g b^{(g)}_{g-1}-\frac{20}{3}b^{(g)}_{g-2} \Bigr];
\label{V44}\\
V^{(g)}_{5,3{-}3}&=&2(g-1)b^{(g)}_{g-1}+\frac{8}{3}b^{(g)}_{g-2};
\label{V5}\\
\sum_{q}q V^{(g)}_{q;3{-}3}&=&gb^{(g)}_{g-1}-\frac{20}{3}b^{(g)}_{g-2}.
\label{qVq}
\eea

Observe that there is another particular combination of $V$'s that produce an interesting relation
\beq
\label{mystery}
(2!)2^2 V^{(g)}_{4,4,3{-}3}+5 V^{(g)}_{5,3{-}3}=(6g-5)(6g-6)V^{(g)}_{3{-}3},\quad\hbox{where}\quad
V^{(g)}_{3{-}3}=\sum_{q=0}^{\mathrm{max}}V^{(g)}_{q;3{-}3}.
\eeq
(We have verified the validity of this relation for $g=3$ using the data from \cite{Mil-Pen}.)

\subsubsection{Blowing up process, $g\to g+1$}

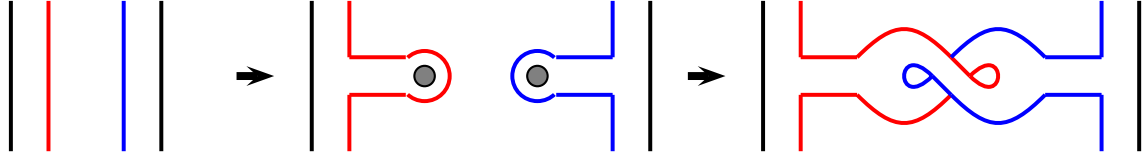
\begin{figure}[h]
{\psset{unit=1}
\begin{pspicture}(-9,-1)(8,1)
\pcline[linewidth=1.5pt](-8,-1)(-8,1)
\pcline[linewidth=1.5pt,linecolor=red](-7.5,-1)(-7.5,1)
\pcline[linewidth=1.5pt,linecolor=blue](-6.5,-1)(-6.5,1)
\pcline[linewidth=1.5pt](-6,-1)(-6,1)
\pcline[linewidth=3pt]{->}(-5,0)(-4.5,0)
\pcline[linewidth=1.5pt](-4,-1)(-4,1)
\pcline[linewidth=1.5pt,linecolor=red](-3.5,-1)(-3.5,-0.25)
\pcline[linewidth=1.5pt,linecolor=red](-3.5,1)(-3.5,0.25)
\pscircle[linewidth=1.5pt,linecolor=red](-2.5,0){.36}
\pswedge[linecolor=white,fillstyle=solid,fillcolor=white](-2.5,0){1}{135}{225}
\pscircle[fillstyle=solid,fillcolor=gray](-2.5,0){.15}
\pcline[linewidth=1.5pt,linecolor=red](-3.5,0.25)(-2.75,0.25)
\pcline[linewidth=1.5pt,linecolor=red](-3.5,-0.25)(-2.75,-0.25)
\rput{180}(-3.5,0){
\pcline[linewidth=1.5pt](-4,-1)(-4,1)
\pcline[linewidth=1.5pt,linecolor=blue](-3.5,-1)(-3.5,-0.25)
\pcline[linewidth=1.5pt,linecolor=blue](-3.5,1)(-3.5,0.25)
\pscircle[linewidth=1.5pt,linecolor=blue](-2.5,0){.36}
\pswedge[linecolor=white,fillstyle=solid,fillcolor=white](-2.5,0){1}{135}{225}
\pscircle[fillstyle=solid,fillcolor=gray](-2.5,0){.15}
\pcline[linewidth=1.5pt,linecolor=blue](-3.5,0.25)(-2.75,0.25)
\pcline[linewidth=1.5pt,linecolor=blue](-3.5,-0.25)(-2.75,-0.25)}
\pcline[linewidth=3pt]{->}(1,0)(1.5,0)
\pcline[linewidth=1.5pt](2,-1)(2,1)
\rput(0.5,0){
\pcline[linewidth=1.5pt,linecolor=red](2,-1)(2,-0.25)
\pcline[linewidth=1.5pt,linecolor=red](2,1)(2,0.25)
\pcline[linewidth=1.5pt,linecolor=red](2,0.25)(2.75,0.25)
\pcline[linewidth=1.5pt,linecolor=red](2,-0.25)(2.75,-0.25)
\psbezier[linewidth=1.5pt,linecolor=red](4.25,0)(4.75,.5)(4.75,-0.5)(4.25,0)
\pcline[linewidth=1.5pt,linecolor=red](4.25,0)(4,0.25)
\psbezier[linewidth=1.5pt,linecolor=red](4,0.25)(3.5,0.75)(3.25,0.75)(2.75,0.25)
\psbezier[linewidth=1.5pt,linecolor=red](4,-0.25)(3.5,-0.75)(3.25,-0.75)(2.75,-0.25)}
\rput{180}(8.5,0){
\pcline[linewidth=1.5pt,linecolor=blue](2,-1)(2,-0.25)
\pcline[linewidth=1.5pt,linecolor=blue](2,1)(2,0.25)
\pcline[linewidth=1.5pt,linecolor=blue](2,0.25)(2.75,0.25)
\pcline[linewidth=1.5pt,linecolor=blue](2,-0.25)(2.75,-0.25)
\psbezier[linewidth=1.5pt,linecolor=blue](4.25,0)(4.75,.5)(4.75,-0.5)(4.25,0)
\pcline[linewidth=1.5pt,linecolor=blue](4.25,0)(4,0.25)
\psbezier[linewidth=1.5pt,linecolor=blue](4,0.25)(3.5,0.75)(3.25,0.75)(2.75,0.25)
\psbezier[linewidth=1.5pt,linecolor=blue](4,-0.25)(3.5,-0.75)(3.25,-0.75)(2.75,-0.25)}
\pcline[linewidth=1.5pt](7,-1)(7,1)
\end{pspicture} }
\caption{\small The procedure of gluing the handle into two sides of two arbitrary edges of a three-valent
graph $\Gamma_{3\dots3}$, which increase the genus by one.
We can think about it as of blowing up a handle from a pair of punctures.}
\label{fi:handle}
\end{figure}

We now consider the ``inverse'' process depicted in
Fig.~\ref{fi:handle}, which enables us to blow up a handle from a
pair of marked sides of edges in a graph from
$\Gamma^{(g)}_{q;3{-}3}$; the number of 2-cycles is irrelevant here.
At the first stage we allow ``bubbling'' as in the middle diagram in
Fig.~\ref{fi:handle} of two sides of edges; we must now allow this
bubbling on the marked edge as well. We also must present this
marked edge as a subdiagram comprising three edges joined in a single
vertex: two edges are incident to the rest of the diagram (their
ends are the ends of the marked edge and are therefore always
different), the third edge is the tail. (We can consider bubbling
process in order, then, on the first stage, we have $2(6g-1)$
possibilities of setting a bubble on an edge side whereas on the
second stage we have already $2(6g+1)$ such possibilities because we
increased the total number of edges by two in the first process. So,
the total number of possibilities is
$$
\frac12 2^2 (6g-1)(6g+1)=2(6g-1)(6g+1).
$$
Every time we bubble a graph from $\Gamma^{(g)}_{3{-}3}$ we obtain a
graph from $\Gamma^{(g+1)}_{q;3{-}3}$ with $q\ne 0$. 
Vice versa, every graph from $\Gamma^{(g+1)}_{q;3{-}3}$ with nonzero $q$ can
be obtained in exactly $q$ ways from graphs from
$\Gamma^{(g)}_{3{-}3}$. (Note that the number of 2-cycles does not
necessarily increase in this process: if we bubble a side of an edge
entering a 2-cycle in the initial graph, we destroy this 2-cycle,
so, in principle, we can even reduce the number of 2-cycles in this
process, but every time we obtain a graph of genus $g+1$ with at
least one 2-cycle.

We therefore have the relation
\beq
\label{Vg->Vg+1}
2(6g-1)(6g+1)V^{(g)}_{3{-}3}=\sum_{q=0}^{\mathrm{max}}qV^{(g+1)}_{q;3{-}3},
\eeq
from which, substituting the result in (\ref{qVq}) and recalling that $V^{(g)}_{3{-}3}$ is merely
$b^{(g)}_{g-1}$, we obtain the new relation on $b$'s:
\beq
\label{bg-bg-1}
2(6g-1)(6g+1)b^{(g)}_{g-1}=(g+1)b^{(g+1)}_{g}-\frac{20}{3}b^{(g+1)}_{g-1},
\eeq
from which we immediately find that
\beq
\label{bgg-3}
b^{(g)}_{g-2}=\frac{3}{10}\frac{g(g-1)}{2g-1}b^{(g)}_{g-1},
\eeq
which coincides with (\ref{bgg-2}).

\subsection*{Conclusion}
Application of topological recursion (TR) to constructing generating 
functions for cohomological field theories is becoming an important issue in 
contemporary mathematical physics (see, e.g., the recent paper \cite{FLZ14} where all genus all descendants equivariant Gromov-Witten invariants of ${\mathbb P}^1$ were constructed using TR). In this respect, it seems interesting to understand the status of 
Givental-type decompositions in the quantum spectral curve approach.

\subsection*{Acknowledgments}
Sections 2, 3, and 6.2 were written by L. O. Chekhov, and Secs. 1, 4, 5, and 7 and also the other parts of Sec. 6
were written by J. E. Andersen, P. Norbury, and R. C. Penner.
The research of L. O. Chekhov was funded by a grant from the Russian Science Foundation (Project No. 14-50-00005) and was performed in Steklov Mathematical Institute of Russian Academy of Sciences.





\end{document}